\documentclass[draft]{agujournal2019}
\usepackage{url} 
\usepackage{lineno}
\usepackage[inline]{trackchanges} 
\usepackage{soul}

\draftfalse

\journalname{JGR: Planets}

\usepackage{pdflscape,epstopdf,textgreek,tabularx,colortbl}
\usepackage[shortlabels]{enumitem}
\newcolumntype{L}{>{\raggedright\arraybackslash}X}
\begin{document}

%
%

\title{Enhanced C\textsubscript{2}H\textsubscript{2} absorption within Jupiter's southern auroral oval from Juno UVS observations}

%
%

\authors{Rohini S. Giles\affil{1}, Vincent Hue\affil{1, 2}, Thomas K. Greathouse\affil{1}, G. Randall Gladstone\affil{1,3}, Joshua A. Kammer\affil{1}, Maarten H. Versteeg\affil{1}, Bertrand Bonfond\affil{4}, Denis C. Grodent\affil{4}, Jean-Claude G\'{e}rard\affil{4}, James A. Sinclair\affil{5}, Scott J. Bolton\affil{1}, and Steven M. Levin\affil{5}}

\affiliation{1}{Space Science and Engineering Division, Southwest Research Institute, San Antonio, Texas, USA}
\affiliation{2}{Aix-Marseille Université, CNRS, CNES, Institut Origines, LAM, Marseille, France}
\affiliation{3}{Department of Physics and Astronomy, University of Texas at San Antonio, San Antonio, Texas, USA}
\affiliation{4}{Laboratoire de Physique Atmosphérique et Planétaire, STAR Institute, Université de Liège, Liège, Belgium}
\affiliation{5}{Jet Propulsion Laboratory, Pasadena, California, USA}

%
%


%
%

\begin{abstract}

Reflected sunlight observations from the Ultraviolet Spectrograph (UVS) on the Juno spacecraft were used to study the distribution of acetylene (C\textsubscript{2}H\textsubscript{2}) at Jupiter's south pole. We find that the shape of the C\textsubscript{2}H\textsubscript{2} absorption feature varies significantly across the polar region, and this can be used to infer spatial variability in the C\textsubscript{2}H\textsubscript{2} abundance. There is a localized region of enhanced C\textsubscript{2}H\textsubscript{2} absorption which coincides with the location of Jupiter's southern polar aurora; the C\textsubscript{2}H\textsubscript{2} abundance poleward of the auroral oval is a factor of 3 higher than adjacent quiescent, non-auroral longitudes. This builds on previous infrared studies which found enhanced C\textsubscript{2}H\textsubscript{2} abundances within the northern auroral oval. This suggests that Jupiter's upper-atmosphere chemistry is being strongly influenced by the influx of charged auroral particles and demonstrates the necessity of developing ion-neutral photochemical models of Jupiter's polar regions.

\end{abstract}

\section*{Plain Language Summary}

The UVS instrument on the Juno mission to Jupiter measures ultraviolet sunlight that is reflected from the planet's upper atmosphere and these observations can be used to measure the abundances of different gases in the stratosphere. In this paper, we study the spatial distribution of the molecule acetylene at Jupiter's south pole. We find that there is a significant increase in the acetylene abundance at the location of Jupiter's southern aurora. This suggests that the charged particles that travel along magnetic field lines towards Jupiter's poles and produce bright auroral emission also have a strong influence on the chemistry of Jupiter's upper atmosphere. 

%
%

\section{Introduction}

Jupiter's strong magnetic field gives rise to the brightest and most energetic auroras in the Solar System. These auroras comprise main auroral emissions, which form a distorted and often discontinuous oval around the magnetic pole; complex polar emissions within the auroral ovals; and diffuse equatorward emissions, including the auroral satellite footprints \cite{grodent15}.

Auroral processes lead to localized heating of the atmosphere within the auroral regions \cite{caldwell80,drossart93b,sinclair17b} and may also be responsible for the high upper atmosphere temperatures across the entire planet \cite{waite83,odonoghue21}. Several previous studies have also suggested that the precipitation of high-energy charged particles may lead to the modification of Jupiter's stratospheric chemistry in the auroral regions. Using infrared observations from Voyager IRIS, \citeA{kim85} observed enhanced emission from a range of hydrocarbons, including C\textsubscript{2}H\textsubscript{2}, C\textsubscript{2}H\textsubscript{4} and C\textsubscript{2}H\textsubscript{6}, within Jupiter's northern auroral oval. They concluded that the enhanced emission was due to increased abundances of these species, although in the infrared spectral region they focus on, there is significant degeneracy between chemical abundances and stratospheric temperatures. \citeA{drossart86} obtained ground-based mid-infrared observations of Jupiter and similarly observed a C\textsubscript{2}H\textsubscript{2} emission hot-spot that coincided with the northern aurora, which they noted could be caused either by an enhancement in the C\textsubscript{2}H\textsubscript{2} abundance or an alteration in the thermal profile. 

More recently, \citeA{sinclair17b,sinclair18,sinclair19} have sought to break the degeneracy between chemistry and temperature in order to map the distribution of hydrocarbons in Jupiter's polar regions. \citeA{sinclair17b} used a radiative transfer model to simultaneously constrain the temperature profile and chemical abundances using Voyager IRIS and Cassini CIRS infrared observations. \citeA{sinclair18} performed a similar analysis using high spectral resolution mid-infrared observations obtained from a ground-based telescope, making use of a two-step retrieval process that first used  CH\textsubscript{4} and H\textsubscript{2} S(1) emission to constrain the temperature and then retrieved the hydrocarbon abundances. These studies found that C\textsubscript{2}H\textsubscript{2} and C\textsubscript{2}H\textsubscript{4} are enhanced within Jupiter's northern auroral oval and \citeA{sinclair17b} suggested that the precipitation of charged particles leads to an increase in ion-neutral and electron recombination reactions which, by analogy with Titan's hydrocarbon chemistry, would favor the production of unsaturated hydrocarbons, including C\textsubscript{2}H\textsubscript{2} \cite{delahaye08}. \citeA{sinclair19} extended this analysis to study additional hydrocarbons and found that C\textsubscript{2}H\textsubscript{4}, CH\textsubscript{3}C\textsubscript{2}H and C\textsubscript{6}H\textsubscript{6} are also enhanced within the auroral region.

In this paper, we build on these previous studies by using ultraviolet reflected sunlight observations from the UVS instrument on the Juno mission to study the distribution of C\textsubscript{2}H\textsubscript{2} in Jupiter's polar regions. C\textsubscript{2}H\textsubscript{2} is the hydrocarbon with the strongest spectral signature in the reflected sunlight portion of the UVS spectrum and ultraviolet observations have the advantage of being insensitive to the thermal structure, and therefore complement the earlier infrared studies. The polar orbit of the Juno spacecraft provides a unique viewpoint of Jupiter's auroral regions, particularly the southern aurora which cannot be easily seen from Earth or from equatorial-orbit spacecraft. In Section~\ref{sec:results} we present reflectance maps of Jupiter's poles and show how the shape of the reflectance spectra varies across the polar region. By comparing reflectance maps made at different wavelength intervals, we can infer the C\textsubscript{2}H\textsubscript{2} distribution and we discuss how these results are impacted by uncertainty in the haze opacity and C\textsubscript{6}H\textsubscript{6} abundance. In Section~\ref{sec:discussion} we compare the proxy C\textsubscript{2}H\textsubscript{2} abundance maps with previous results obtained using infrared observations and discuss the possible mechanisms that drive the C\textsubscript{2}H\textsubscript{2} distribution.

%
%

\section{Juno UVS observations}
\label{sec:observations}

The Ultraviolet Spectrograph \cite<UVS,>{gladstone17} is a far-ultraviolet imaging spectrograph on NASA's Juno mission, which has been in a highly elliptical polar orbit around Jupiter since July 2016 \cite{Bolton17}. UVS covers the 68-210 nm spectral range with a spectral resolution that varies between 1.3 nm and 3.0 nm depending on the position along the instrument's slit \cite{greathouse13}. The primary scientific goal of the UVS instrument is to characterize Jupiter's far-ultraviolet auroral emissions and this spectral range was selected to include the H Lyman series and the Lyman, Werner, and Rydberg band systems of H\textsubscript{2}. However, the longer wavelengths covered by UVS extend into a spectral region that is dominated by reflected sunlight and this part of the spectrum is sensitive to hydrocarbons and aerosols \cite{gladstone17}. In this paper, we focus on the 175-205 nm reflected sunlight segment of the UVS spectrum. 

Juno is a spin-stabilized spacecraft with a rotation period of $\sim$30 seconds. The UVS instrument slit is nominally parallel to the spacecraft's spin axis, allowing the slit to sweep across a swath of planet as the spacecraft rotates. In this paper, we exclusively use data from the wide part of the dog-bone shaped slit, as it has a better signal-to-noise ratio, particularly at the longer wavelengths where the instrument detector is less sensitive \cite{hue19b, hue21b}. UVS data is recorded in a pixel-list time-tagged format; by considering the position along the slit and the spin phase of the spacecraft at the time of observation, each photon can be assigned to a latitude and longitude on the planet. This geometric information can then be used to produce spatial maps of the ultraviolet radiation \cite{gladstone17}. In-flight radiometric calibration is achieved by observation of UV-bright O, B, A stars \cite{hue19b}, and radiances can be converted to reflectances by dividing by the solar spectrum \cite{woods09}, corrected for the solar incidence angle.

UVS obtains observations of Jupiter for several hours on either side of each perijove (the point of closest approach for the spacecraft), covering both the approach over Jupiter's north pole and the subsequent departure over the south pole. When the spacecraft passes through Jupiter's radiation belts, the UVS instrument records high background noise from high energy electrons and ions, and so data acquisition is temporarily paused to both protect the instrument from enhanced degradation \cite{kammer19} and to limit the volume of low-quality noisy data. Due to the precession of the Juno orbit over the course of the mission, the location of the perijove has gradually moved northwards over time, which has moved the approach observations further into the radiation belts. This means that the amount of UVS data obtained from the northern polar region has decreased over time \cite{gladstone19}  and when we co-add reflected sunlight data from many perijoves, as described in Section~\ref{sec:maps}, the signal-to-noise ratio and spatial coverage is significantly better for the south pole. 

\section{Analysis and Results}
\label{sec:results}

\subsection{Molecular absorption}
\label{sec:molecules}

\begin{figure}
\centering
\includegraphics[width=9cm]{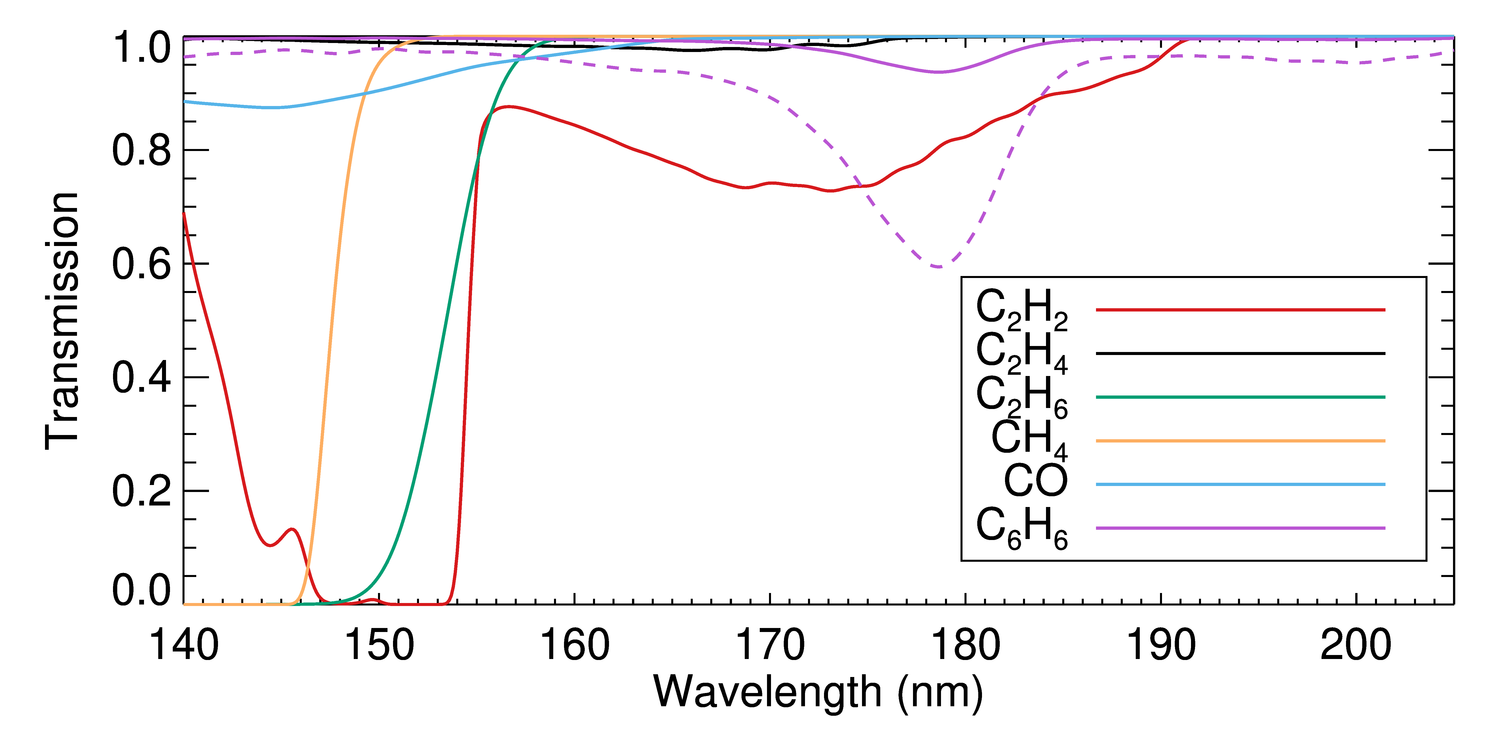}
\caption{Absorption spectra for six molecules with significant absorption in the 140--205 nm spectral range. The transmission is defined as the transmission between an altitude of 100 km (5 mbar) and the top of the atmosphere. The absorption cross-sections have been smoothed to match the spectral resolution of the UVS wide slit. Two abundances are shown for C\textsubscript{6}H\textsubscript{6}: the a priori value (solid line) and 8$\times$ the a priori value (dashed line).}
\label{fig:gases}
\end{figure}

Figure~\ref{fig:gases} shows the absorption spectrum for C\textsubscript{2}H\textsubscript{2}, as well as five other molecular species in Jupiter's stratosphere that produce non-negligible absorption in the 140--205 nm spectral range. The transmission is defined as the transmission between an altitude of 100 km (5 mbar) and the top of the atmosphere. The column density for each gas was calculated using the atmospheric composition model described by Model C in \citeA{moses05}, with the exception of C\textsubscript{6}H\textsubscript{6}. As noted in~\citeA{sinclair19}, the peak of the C\textsubscript{6}H\textsubscript{6} distribution is likely to be higher in the atmosphere than predicted in \citeA{moses05} due to additional ion sources. We therefore follow a similar approach to~\citeA{sinclair19} and use an a priori vertical profile for C\textsubscript{6}H\textsubscript{6} that is 0.1\% of the a priori vertical profile for C\textsubscript{2}H\textsubscript{2}; this C\textsubscript{6}H\textsubscript{6} a priori vertical profile matches the \citeA{moses05} Model C C\textsubscript{6}H\textsubscript{6} profile at 3--5 mbar but has significantly higher densities at higher altitudes. Absorption cross-sections were obtained from \citeA{keller-rudek13} and were smoothed to match the 2.2 nm spectral resolution of the UVS wide slit.

Figure~\ref{fig:gases} shows that while there are multiple highly absorbing species at \textless160 nm, C\textsubscript{2}H\textsubscript{2} and C\textsubscript{6}H\textsubscript{6} are the only species with non-negligible absorption at \textgreater175 nm. C\textsubscript{2}H\textsubscript{2} has a broad absorption feature that has a local maximum at $\sim$172 nm, while C\textsubscript{6}H\textsubscript{6} has a narrower feature that peaks at $\sim$179 nm. The solid purple line shows the absorption produced by the a priori C\textsubscript{6}H\textsubscript{6} distribution, which is considerably weaker than the C\textsubscript{2}H\textsubscript{2} absorption. However, the C\textsubscript{6}H\textsubscript{6} abundance is highly uncertain, and \citeA{sinclair19} showed that within the northern auroral oval the abundance can be as high as 0.8\% of the C\textsubscript{2}H\textsubscript{2} abundance, a factor of 8 higher than our a priori. The transmission curve for this higher value is shown by the dashed purple line in Figure~\ref{fig:gases}, showing that C\textsubscript{6}H\textsubscript{6} can also have a significant impact on the shape of Jupiter's 175--205 nm spectrum and must therefore be considered in the following analysis.

\subsection{Reflectance maps}
\label{sec:maps}

\begin{figure}
\centering
\includegraphics[width=12cm]{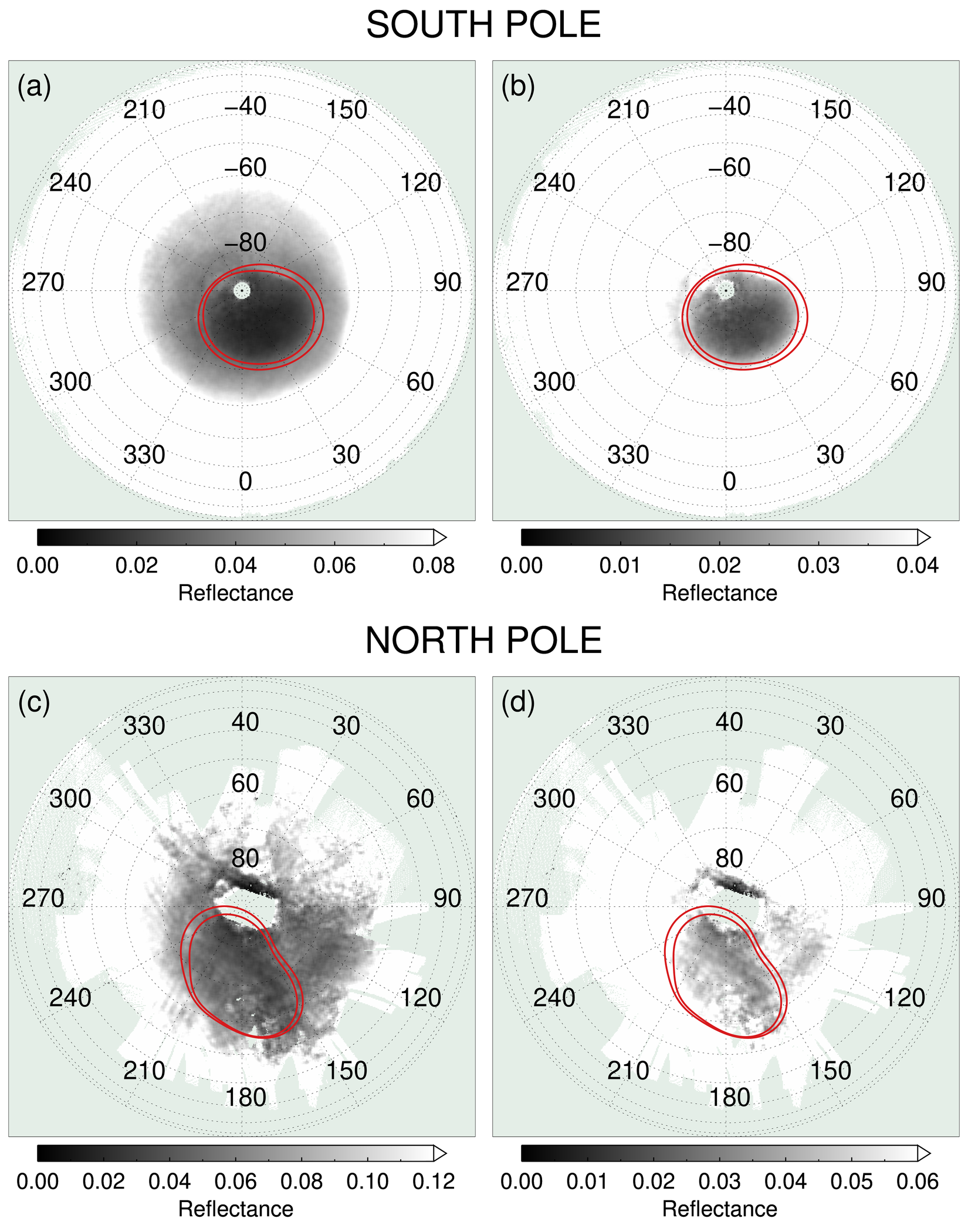}
\caption{UV reflectance maps of Jupiter's south pole ((a) and (b)) and north pole ((c) and (d)). The figures show the mean reflectance in the 175--190 nm spectral range, which is sensitive to C\textsubscript{2}H\textsubscript{2} absorption. (b) and (d) show the same data as (a) and (c) but have a different grayscale stretch to highlight different morphologies. The average locations of the expanded and contracted auroral ovals \cite{bonfond17b} are shown in red.}
\label{fig:maps}
\end{figure}

Figure~\ref{fig:maps} shows reflectance maps of Jupiter's southern and northern polar regions at wavelengths of 175--190 nm. As described in Section~\ref{sec:observations}, Juno UVS observations were converted from radiance to reflectance by correcting for the solar incidence angle and dividing by the solar spectrum \cite{woods09}. The maps shown in Figure~\ref{fig:maps} combine all dayside (local time of 8 AM to 4 PM, solar zenith angle less than 85$^{\circ}$) data obtained during the first 46 perijoves of the Juno mission (July 2016 - November 2022). The maps comprise observations made at a range of different spacecraft positions, spanning radial distances of 1.3--9.7 R\textsubscript{J}.

Because of the precession of the spacecraft orbit, the southern hemisphere (Figures~\ref{fig:maps}(a) and~\ref{fig:maps}(b)) has significantly better spatial coverage and signal-to-noise ratios than the northern hemisphere (Figures~\ref{fig:maps}(c) and~\ref{fig:maps}(d)). The higher noise level means that some `striping' is visible in the northern data. Figures in the right column are identical to figures in the left column but have a different grayscale stretch to highlight different morphologies. The average locations of the expanded and contracted auroral ovals \cite{bonfond17b} are shown in red.

The wavelength range of 175--190 nm was chosen because this is a spectral region that is sensitive to absorption by C\textsubscript{2}H\textsubscript{2}, as shown in Figure~\ref{fig:gases}. While slightly shorter wavelengths would be even more sensitive (C\textsubscript{2}H\textsubscript{2} has a local absorption maximum at $\sim$172 nm), shorter wavelengths are also increasingly sensitive to H\textsubscript{2 }auroral emissions that peak near 160 nm and lead to misleadingly high reflectances along the auroral ovals; 175--190 nm represents a spectral region that is sensitive to C\textsubscript{2}H\textsubscript{2} while minimizing the contamination from auroral emission.

Figure~\ref{fig:maps}(a) shows that the reflectance at all longitudes decreases towards the south pole. This is a well-known phenomenon and is caused by presence of stratospheric hazes at both poles; these hazes lead to a distinctive appearance in both the infrared and the ultraviolet \cite{west04}. In the infrared, there is a bright  ``polar hood'' in methane-band observations, due to the fact that the stratospheric haze particles scatter light before it has the chance to be absorbed by methane. In contrast, the UV polar hood is dark, as seen in Figure~\ref{fig:maps}(a), because the haze particles are less reflective than the Rayleigh-scattering molecular atmosphere. 

Although the entire polar region in Figure~\ref{fig:maps}(a) is relatively dark, the segment within the southern auroral oval (marked by the red lines) is darker than the adjacent longitudes. This can be seen more clearly in Figure~\ref{fig:maps}(b), which presents the same data as Figure~\ref{fig:maps}(a) but with a different grayscale stretch to highlight morphology within the darkest regions. There is a distinct dark patch which aligns closely with the location of the auroral oval; the reflectance within the auroral oval is 0.02, while the mean reflectance at similar latitudes outside the oval is 0.05. A similar effect can be seen at the north pole, shown in Figures~\ref{fig:maps}(c) and~\ref{fig:maps}(d). These observations are noisier and have decreased coverage, but the same approximate trend can be seen; the region within the auroral oval has an even lower reflectance than the rest of the pole. These observations could be due to an increase in the haze concentration within the auroral ovals, which would lead to a local enhancement of the effect that causes the broader polar hood. Alternatively, it could be due to an increase in the C\textsubscript{2}H\textsubscript{2} and/or C\textsubscript{6}H\textsubscript{6} abundances, both of which would lead to increased absorption at 175--190 nm and therefore lower reflectance. In order to distinguish between these scenarios, we consider the full UVS spectra in Section~\ref{sec:spectra}.

\subsection{Spectra}
\label{sec:spectra}

To further explore how the UV reflectance varies across the polar region, we now consider the full UVS spectra. In order to obtain a sufficient signal-to-noise ratio, we focus solely on the southern polar region and co-add data to produce two representative spectra: (i) the polar auroral region, defined as within the contracted auroral oval, and (ii) the polar quiescent region, defined as outside the expanded auroral oval, but still poleward of 67.2$^{\circ}$S (the most-equatorward point of the expanded oval). In both cases, the data is limited to local times of 8 AM to 4 PM and is corrected for the solar incidence angle, as in Figure~\ref{fig:maps}.

\begin{figure}
\centering
\includegraphics[width=14cm]{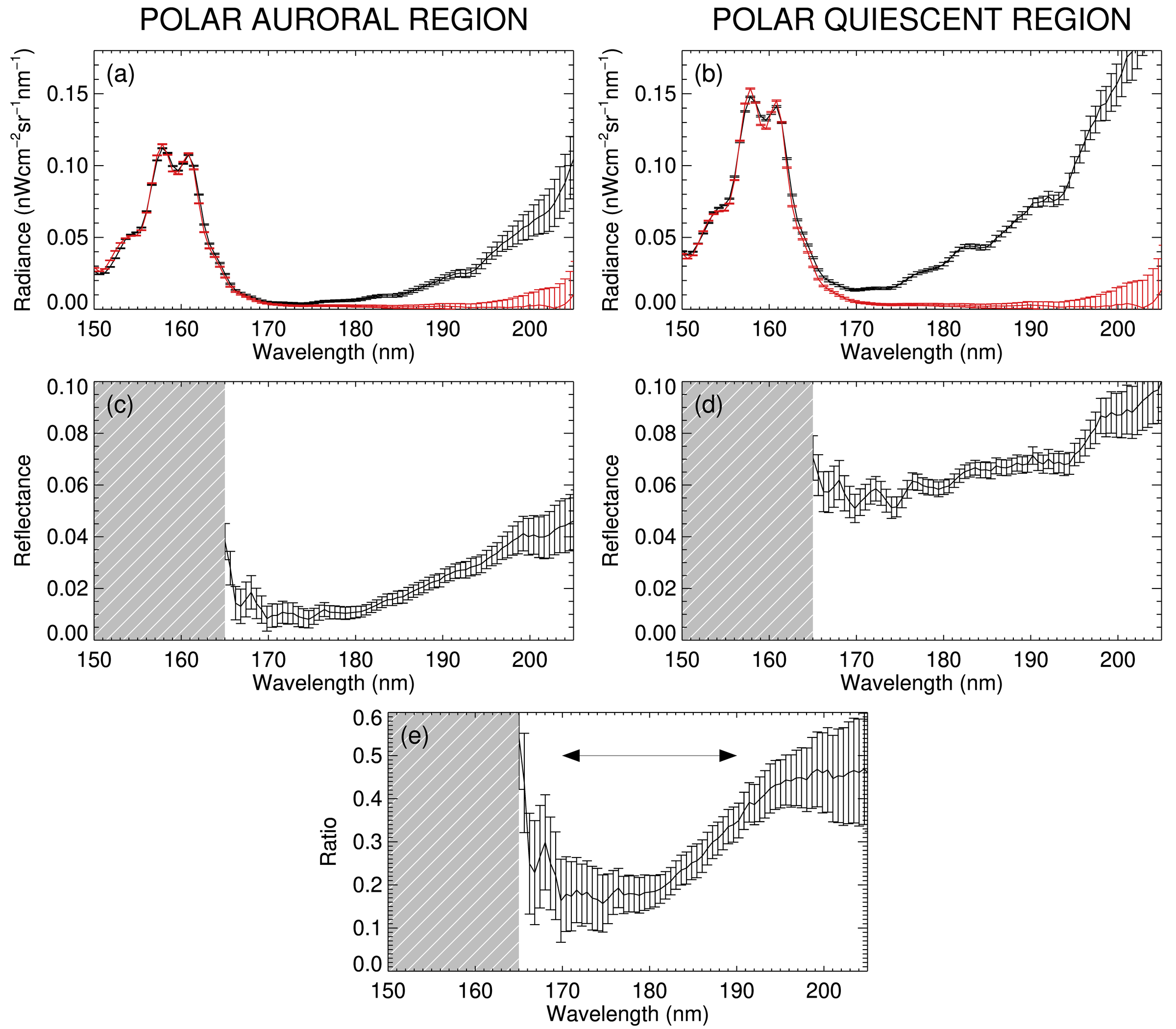}
\caption{Comparison of UV spectra from the polar auroral and quiescent regions of Jupiter's southern pole. The polar auroral region is defined as within the contracted auroral oval and the polar quiescent region is defined as outside the expanded auroral oval but poleward of  of 67.2$^{\circ}$S. (a) and (b) show dayside spectra from the two regions (black) alongside a pure auroral emission spectrum (red). (c) and (d) show the reflectance spectra from the two regions once the auroral emission component has been removed; the spectral range dominated by the auroral emission is shaded out. (e) shows the ratio of the two spectra shown in (c) and (d). The wavelength range where an absorption feature is present is highlighted by the arrows.}
\label{fig:spectra}
\end{figure}

The representative spectra from inside and outside the southern auroral oval are shown in black in Figures~\ref{fig:spectra}(a) and~\ref{fig:spectra}(b) respectively. Because these spectra are obtained from the dayside of the planet in a spatial region with significant auroral emission, they consist of a combination of reflected sunlight and auroral emission. At longer wavelengths (\textgreater175 nm), the reflected sunlight component dominates, and at shorter wavelengths (\textless165 nm), the auroral emission dominates. The spectrum from the polar `quiescent' region still has a strong auroral emission component due to the diffuse equatorward emissions from the regions closest to the auroral oval. 

In order to obtain a spectrum that consists of purely reflected sunlight, we first obtained a pure auroral spectrum by coadding polar data obtained during night time. This pure auroral spectrum is shown in red in Figures~\ref{fig:spectra}(a) and~\ref{fig:spectra}(b), and in each case, has been scaled to match the average radiance of the dayside spectrum at 157--160 nm. By subtracting these auroral spectra from the dayside spectra, and then dividing by the solar spectrum, we were able to produce the `clean' reflectance spectra shown in Figures~\ref{fig:spectra}(c) and~\ref{fig:spectra}(d). 

Figures~\ref{fig:spectra}(c) and~\ref{fig:spectra}(d) have noticeably different spectral shapes. Figure~\ref{fig:spectra}(c), the reflectance spectrum from the polar auroral region, has a steep slope between 175 nm and 195 nm, while Figure~\ref{fig:spectra}(d), the spectrum from the quiescent region, is significantly flatter. Figure~\ref{fig:spectra}(e) further compares the two reflectance spectra by showing a ratio of the polar auroral region to the quiescent region. At 195--205 nm, the relative reflectance from the polar auroral region is relatively constant and $\sim$50\% of the reflectance at quiescent longitudes. This then decreases to $\sim$20\% by 175 nm, before increasing again at shorter wavelengths. The shape shown in Figure~\ref{fig:spectra}(e) is very similar to the shape of the C\textsubscript{2}H\textsubscript{2} absorption feature shown in Figure~\ref{fig:gases}, and is also consistent with the presence of C\textsubscript{6}H\textsubscript{6}. This suggests that the decreased reflectance at 175--190 nm in the polar auroral region is at least partially driven by an increase in these hydrocarbons.

\subsection{Proxy C\textsubscript{2}H\textsubscript{2} and C\textsubscript{6}H\textsubscript{6} abundance maps}
\label{sec:modeling}

In \citeA{giles21b}, we used a radiative transfer and retrieval algorithm to model Juno UVS reflected sunlight spectra from Jupiter's low- and mid-latitudes in order to retrieve the zonally-averaged C\textsubscript{2}H\textsubscript{2} abundances. Ideally, we would perform a similar analysis for the polar region, in order to produce spatially-resolved C\textsubscript{2}H\textsubscript{2} maps and to determine the extent to which any variability in the C\textsubscript{6}H\textsubscript{6} abundance may affect these results. However, the high solar incidence angle at the poles means that the signal-to-noise ratios of the reflected sunlight spectra are significantly lower than they were at low latitudes, and this is compounded by the fact that we are interested in spatial variability on relatively small scales and therefore cannot co-add large spatial regions to reduce the noise. Instead, we can use the ratio between the reflectance at 175--190 nm and the reflectance at 190--205 nm as a proxy for the C\textsubscript{2}H\textsubscript{2} and C\textsubscript{6}H\textsubscript{6} abundances.

\begin{figure}
\centering
\includegraphics[width=9cm]{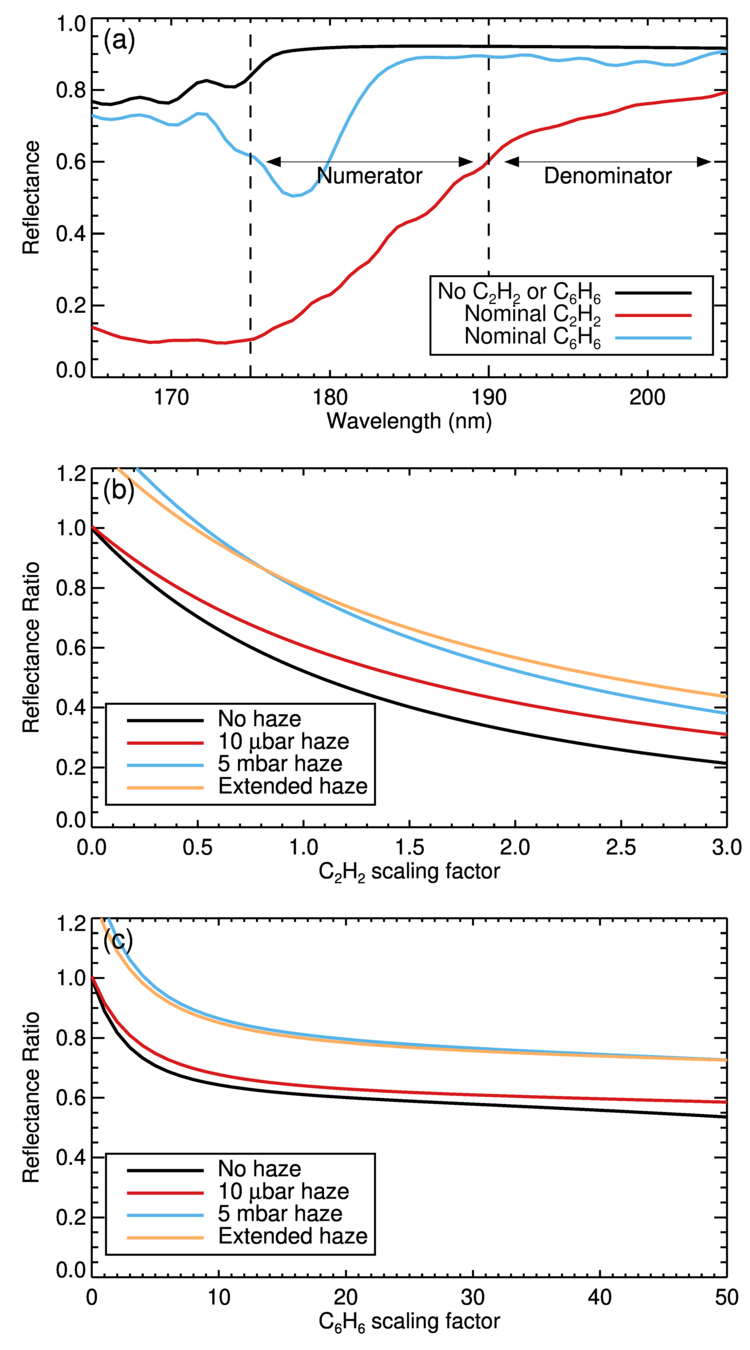}
\caption{(a) Synthetic reflectance spectra of Jupiter assuming a solar incidence angle of 80$^{\circ}$ and a haze-free atmosphere. The spectrum without either C\textsubscript{2}H\textsubscript{2} or C\textsubscript{6}H\textsubscript{6} is shown in black, the spectrum with the nominal C\textsubscript{2}H\textsubscript{2} abundance is shown in red and the spectrum with the nominal C\textsubscript{6}H\textsubscript{6} abundance is shown in blue. The two spectral regions used to calculate the reflectance ratio in (b) and (c) are shown by the arrows. (b) The relationship between the the C\textsubscript{2}H\textsubscript{2} scaling factor relative to the nominal vertical profile and the reflectance ratio for four different cloud models. (c) The relationship between the the C\textsubscript{6}H\textsubscript{6} scaling factor relative to the nominal vertical profile and the reflectance ratio for four different cloud models.}
\label{fig:ratio}
\end{figure}

Figure~\ref{fig:ratio}(a) shows three synthetic reflectance spectra generated using the NEMESIS radiative transfer code \cite{irwin08} and smoothed to match the spectral resolution of Juno UVS. The adaptation of NEMESIS for use in the UV spectral region is described in \citeA{melin20} and \citeA{giles21b}. As described in Section~\ref{sec:molecules}, we use the Model C atmosphere from \citeA{moses05} as the nominal atmospheric model for Jupiter. The black line in Figure~\ref{fig:ratio}(a) shows the reflectance spectrum of Jupiter if C\textsubscript{2}H\textsubscript{2} and C\textsubscript{6}H\textsubscript{6} are removed from the atmosphere, in the absence of any hazes and assuming a solar incidence angle of 80$^{\circ}$. The red line shows how the spectrum changes when the nominal C\textsubscript{2}H\textsubscript{2} abundance is included and the blue line shows the spectrum when the nominal C\textsubscript{6}H\textsubscript{6} abundance (0.1\% of the C\textsubscript{2}H\textsubscript{2} abundance) is included. Both species can clearly have an impact on the ratio between the mean reflectance at 175--190 nm and the mean reflectance at 190--205 nm; the ratio with no C\textsubscript{2}H\textsubscript{2} or C\textsubscript{6}H\textsubscript{6} is 1.00, the ratio with C\textsubscript{2}H\textsubscript{2} alone is 0.52 and the ratio with the C\textsubscript{6}H\textsubscript{6} alone is 0.89. 

Similar reflectance spectra were generated for a range of C\textsubscript{2}H\textsubscript{2} vertical profiles, each consisting of the nominal vertical profile multiplied by a scaling factor between 0 and 3. For each C\textsubscript{2}H\textsubscript{2} scaling factor, the ratio between the mean reflectance at 175--190 nm and the mean reflectance at 190--205 nm was measured. The relationship between the C\textsubscript{2}H\textsubscript{2} scaling factor and the reflectance ratio is shown by the black curve in Figure~\ref{fig:ratio}(b); as the C\textsubscript{2}H\textsubscript{2} abundance increases, the reflectance at 175--190 nm decreases relative to the reflectance at 190--205 nm and so the reflectance ratio decreases. 

The presence of polar hazes can impact the spectral shape. Jupiter's polar hazes are thought to include a range of particle sizes, with mean radii of approximately 0.09--1.1 \textmu m \cite{wong03}. This distribution of particle sizes leads to absorption that is expected to be approximately spectrally flat over the 165--205 nm wavelength range. However, even a spectrally-flat haze can affect the shape of the spectrum if it is located within the altitude region of peak sensitivity; as discussed in \citeA{giles21b}, the longer wavelength end of the UVS spectrum probes deeper into the atmosphere and therefore can be more impacted by the presence of haze, leading to a tilt in the spectrum and increasing the reflectance ratio.

This can be seen by the colored lines in Figure~\ref{fig:ratio}(b) which show the C\textsubscript{2}H\textsubscript{2}--reflectance ratio curves for the three different haze models used in \citeA{giles21b}: a compact layer at 10 \textmu bar, a compact layer at 5 mbar and an extended haze with constant density in the 100--1 mbar range, based on \citeA{wong03}. In each case, the optical thickness of the haze layer was scaled to match the optical thickness retrieved at a latitude of 75$^{\circ}$S in \citeA{giles21b}. For a given C\textsubscript{2}H\textsubscript{2} abundance, the presence of haze acts to increase the reflectance ratio of the synthetic spectrum and for a given observed reflectance ratio, the black no-haze curve shows the corresponding minimum C\textsubscript{2}H\textsubscript{2} scaling factor. However, despite the quantitative differences, the results from all four models shown in Figure~\ref{fig:ratio}(b) are qualitatively similar, showing that as the C\textsubscript{2}H\textsubscript{2} scaling factor increases, the reflectance ratio decreases.

The same method was used to study the effect of varying the C\textsubscript{6}H\textsubscript{6} abundance on the reflectance ratio. Figure~\ref{fig:ratio}(c) shows the relationship between the C\textsubscript{6}H\textsubscript{6} scaling factor and the reflectance ratio for each of the same four haze models. We present a larger range of scaling factors for C\textsubscript{6}H\textsubscript{6} (0--50) than we did for C\textsubscript{2}H\textsubscript{2} (0--3) for two reasons; firstly, the a priori C\textsubscript{6}H\textsubscript{6} abundance is not well constrained, and secondly, a much higher scaling factor is required to significantly decrease the reflectance ratio. Figure~\ref{fig:ratio}(c) shows that a scaling factor of 8, which is the abundance found by \citeA{sinclair19} at Jupiter's northern auroral oval, produces a reflectance ratio of 0.88 using the most realistic extended haze model. The reflectance ratio begins to level off beyond this, and even a scaling factor of 50 can only achieve a reflectance ratio of 0.73. Continuing to increase the C\textsubscript{6}H\textsubscript{6} abundance even further ultimately increases the reflectance ratio instead, as the small C\textsubscript{6}H\textsubscript{6} absorption features at $\sim$200 nm begin to have a greater effect; for the extended haze model, the minimum reflectance ratio that can be achieved using C\textsubscript{6}H\textsubscript{6} alone is 0.61, with a scaling factor of 225$\times$ the nominal value. This shows that while C\textsubscript{6}H\textsubscript{6} does influence the reflectance spectrum 175--205 nm, C\textsubscript{2}H\textsubscript{2} absorption must be primarily responsible for any regions of the planet with very low reflectance ratios.

\begin{figure}
\centering
\includegraphics[width=12cm]{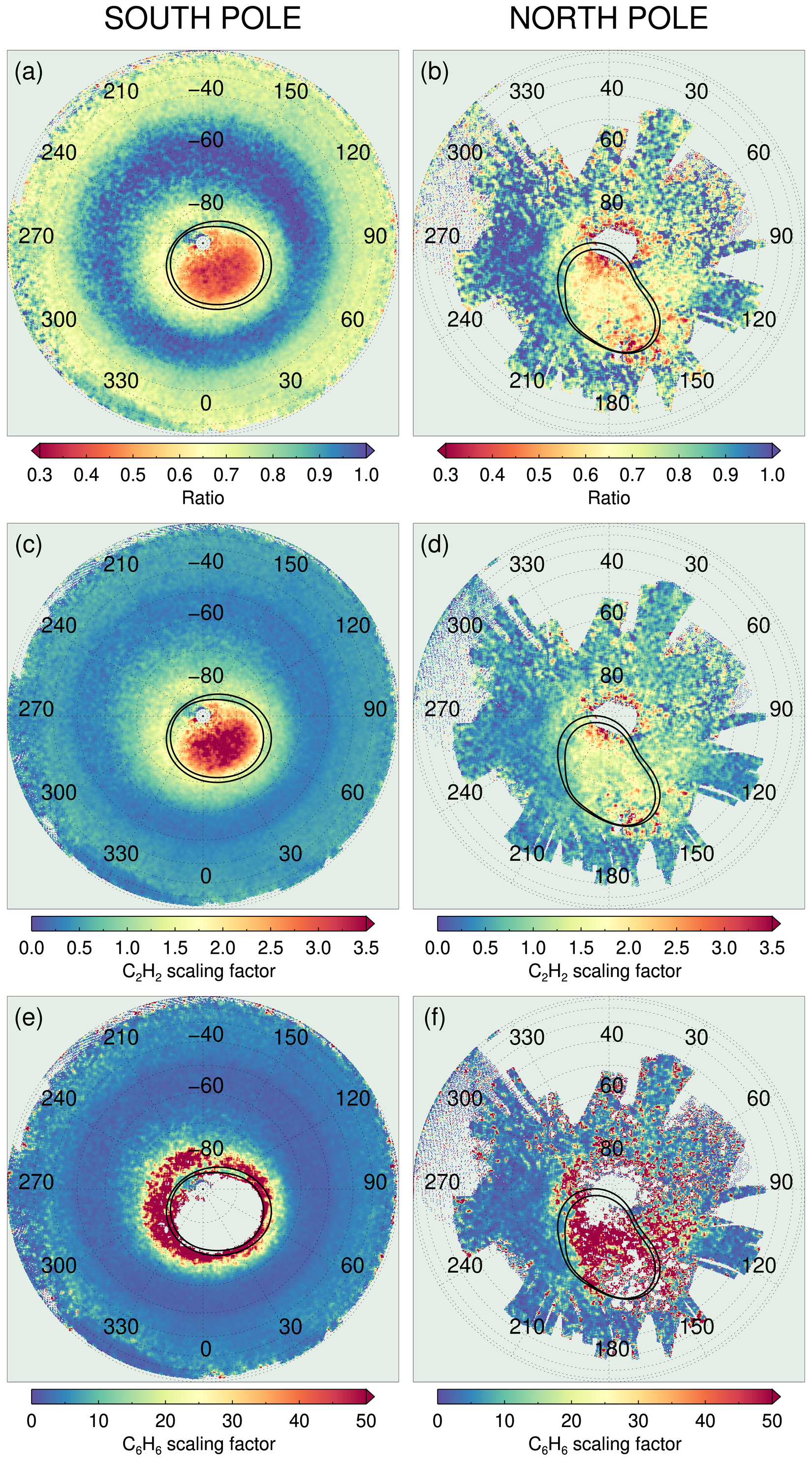}
\caption{(a)--(b) The ratio of the mean reflectance at 175--190 nm to the mean reflectance at 190--205 nm for Jupiter's south and north poles. (c)--(d) Proxy C\textsubscript{2}H\textsubscript{2} abundance maps assuming no C\textsubscript{6}H\textsubscript{6} absorption. (e)--(f) Proxy C\textsubscript{6}H\textsubscript{6} abundance maps assuming no C\textsubscript{2}H\textsubscript{2} absorption. The average locations of the expanded and contracted auroral ovals \cite{bonfond17b} are shown in black.}
\label{fig:ratio_map}
\end{figure}

Figures~\ref{fig:ratio_map}(a) and ~\ref{fig:ratio_map}(b) show reflectance ratio maps of Jupiter's north and south poles. The 175--190 nm reflectance maps shown in Figure~\ref{fig:maps} were divided by equivalent maps covering the 190--205 nm spectral range. As expected from Figure~\ref{fig:spectra}, Figures~\ref{fig:ratio_map}(a) shows that there is a region of low reflectance ratio which coincides with the polar aurora within the southern auroral oval. This low reflectance ratio within the auroral oval is markedly different from the adjacent, quiescent longitudes. Figure~\ref{fig:ratio_map}(b) shows a hint of a similar trend, with a lower reflectance ratio within the northern auroral oval, but the decreased spatial coverage and noisier data make this less clear-cut.

Figure~\ref{fig:ratio_map}(c) shows the southern reflectance ratio map converted into a C\textsubscript{2}H\textsubscript{2} scaling factor, in the absence of C\textsubscript{6}H\textsubscript{6}. For each latitude, we use the appropriate haze opacity retrieved in~\citeA{giles21b} using the extended haze model. The haze retrievals do not extend beyond $\pm75^{\circ}$, so poleward of this a constant haze opacity is assumed. For these polar latitudes, the conversion relationship between reflectance ratio and C\textsubscript{2}H\textsubscript{2} scaling factor is shown by the orange line in Figure~\ref{fig:ratio}(b); for regions of the planet equatorward of $\pm75^{\circ}$, the conversion relationship falls between the orange line and black, haze-free line. As expected from~\citeA{giles21b}, Figure~\ref{fig:ratio_map}(c) shows that outside of the auroral region, the C\textsubscript{2}H\textsubscript{2} abundance decreases towards higher latitudes. In the polar auroral region, however, this trend is reversed. Figure~\ref{fig:ratio_map}(c) shows that the low reflectance within the auroral oval can be explained by a significant enhancement in the C\textsubscript{2}H\textsubscript{2} abundance; the mean C\textsubscript{2}H\textsubscript{2} scaling factor for the polar auroral region within the auroral oval is 2.5 compared to 0.8 at comparable latitudes outside of the auroral oval. Just as  Figure~\ref{fig:ratio_map}(b) showed a hint of the same reflectance ratio trend as Figure~\ref{fig:ratio_map}(a), Figure~\ref{fig:ratio_map}(d) shows a hint of the same C\textsubscript{2}H\textsubscript{2} enhancement within the northern auroral oval, but the noisy data prevents any firm conclusions from being drawn. 

\begin{figure}
\centering
\includegraphics[width=7cm]{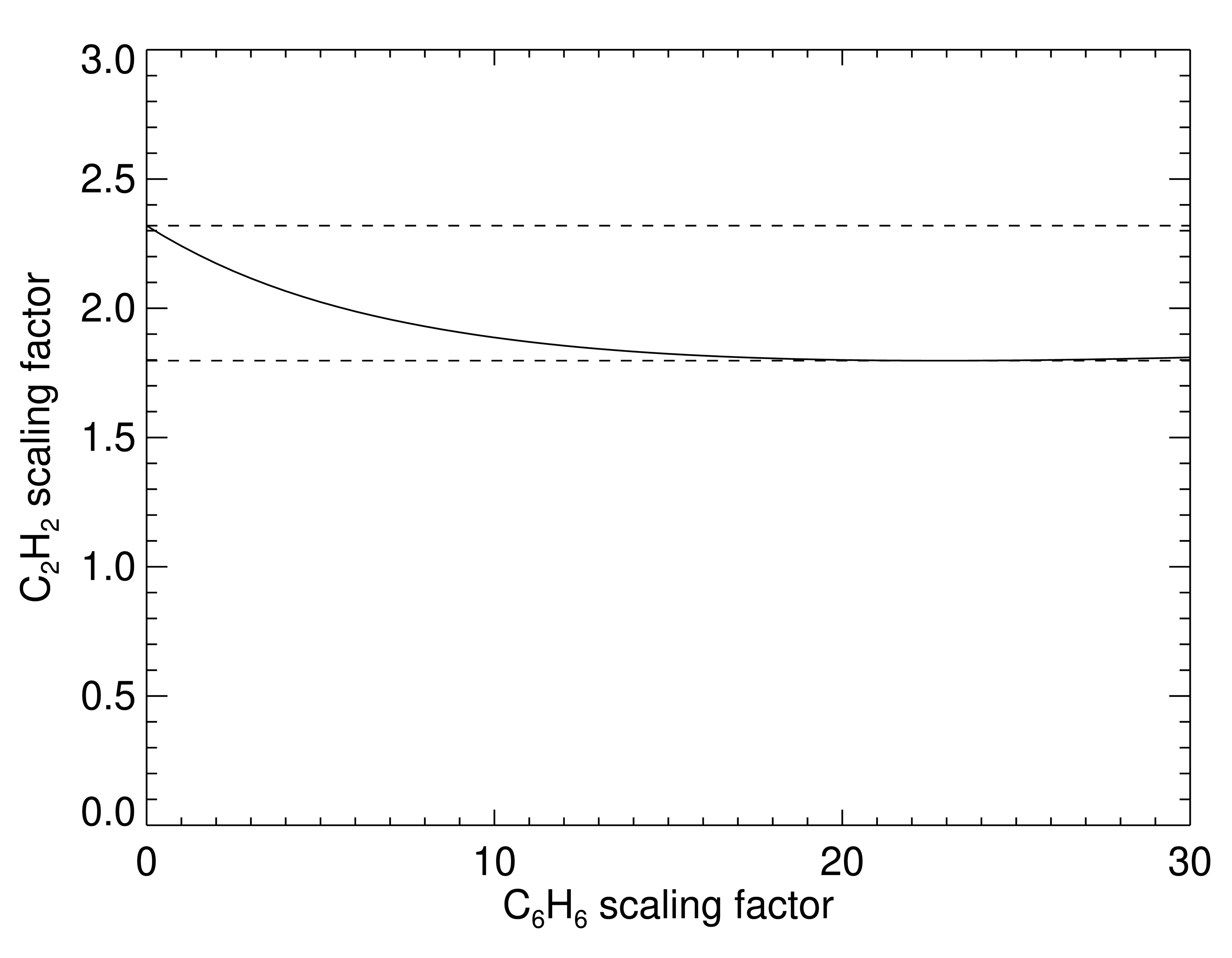}
\caption{The combination of C\textsubscript{2}H\textsubscript{2} and C\textsubscript{6}H\textsubscript{6} scaling factors that are able to reproduce a reflectance ratio of 0.52, the mean value within the southern auroral oval. The minimum and maximum C\textsubscript{2}H\textsubscript{2} scaling factors are marked by the dashed lines.}
\label{fig:c6h6_c2h2}
\end{figure}

Figures~\ref{fig:ratio_map}(e) and~\ref{fig:ratio_map}(f) shows the reflectance ratio maps converted into C\textsubscript{6}H\textsubscript{6} scaling factors instead, in the absence of C\textsubscript{2}H\textsubscript{2}. Figure~\ref{fig:ratio_map}(e) shows that outside of the auroral oval, absorption from C\textsubscript{6}H\textsubscript{6} is technically able to reproduce the observed reflectance ratios, although it should be noted that we know from the full radiative transfer retrievals in~\citeA{giles21b} that the low-latitude spectral shape is actually due to C\textsubscript{2}H\textsubscript{2} absorption. However, within the southern auroral oval, there is no amount of C\textsubscript{6}H\textsubscript{6} that can reproduce the low reflectance ratios observed, which is why that region of the map in Figures~\ref{fig:ratio_map}(e) is empty. This can be further seen in Figure~\ref{fig:c6h6_c2h2}, which shows the combination of C\textsubscript{2}H\textsubscript{2} and C\textsubscript{6}H\textsubscript{6} scaling factors that are together able to produce a synthetic spectrum reflectance ratio of 0.52, the mean value within the southern auroral oval. As the C\textsubscript{6}H\textsubscript{6} increases, slightly less C\textsubscript{2}H\textsubscript{2} is required, but an extremely high levels of C\textsubscript{6}H\textsubscript{6} (\textgreater23$\times$ the a priori) the corresponding C\textsubscript{2}H\textsubscript{2} abundance starts to increase again. For this specific reflectance ratio, the C\textsubscript{2}H\textsubscript{2} scaling factor must fall in the 1.8--2.3 range, marked by the dashed lines. Since C\textsubscript{6}H\textsubscript{6} is unable to reproduce the low reflectance ratios within Jupiter's southern auroral oval, we can conclude that this effect must be primarily driven by a C\textsubscript{2}H\textsubscript{2} enhancement, with any possible enhancement in C\textsubscript{6}H\textsubscript{6} decreasing the C\textsubscript{2}H\textsubscript{2} abundances by up to 20\%. 

It should also be noted that there is some error associated with the haze opacity. In Sections~\ref{sec:maps} and~\ref{sec:spectra}, we noted that the decreased reflectance within the auroral oval could be partially caused by an increase in the haze opacity, as well as an increase in the hydrocarbon abundances, but in Figure~\ref{fig:ratio_map} we assume a constant haze opacity poleward of $\pm75^{\circ}$. However, if the hazes were more opaque within the polar auroral region, the contrast in the C\textsubscript{2}H\textsubscript{2} scaling factor within the oval relative to the scaling factor at quiescent longitudes would actually be even greater than what is shown in Figures~\ref{fig:ratio_map}(c) and~\ref{fig:ratio_map}(d). Figure~\ref{fig:ratio}(b) showed that when additional hazes are present in a model, an even higher C\textsubscript{2}H\textsubscript{2} abundance is required to produce a given reflectance ratio. The presence of additional hazes can therefore only intensify the polar auroral phenomenon shown in Figure~\ref{fig:ratio_map}. 

A similar effect exists with respect to the auroral emissions within the polar region. The wavelength interval of 175--190 nm was chosen in order to minimize the contribution of the auroral emissions which peak at $\sim$160 nm (see Figures~\ref{fig:spectra}(a) and \ref{fig:spectra}(b)). However, the edge of this emission feature may still contribute a small amount of flux, which would act to slightly increase the apparent reflectance within this interval, and therefore increase the ratio of the 175--190 nm reflectance to the 190--205 nm reflectance. Accounting for any such emission component would therefore further decrease the reflectance ratio within and in the immediate vicinity of the auroral ovals, while the reflectance ratio in the quiescent auroral regions would remain the same. Accounting for any residual auroral emission would therefore only act to intensify the polar auroral phenomenon shown in Figure~\ref{fig:ratio_map}.

In summary, we can conclude that the C\textsubscript{2}H\textsubscript{2} abundance is significantly enhanced within Jupiter's southern auroral oval. There are three factors that can act to modify the C\textsubscript{2}H\textsubscript{2} scaling factors presented in Figure~\ref{fig:ratio}(c): the presence of C\textsubscript{6}H\textsubscript{6}, the presence of additional hazes and residual auroral emissions. While the first of these can decrease the auroral C\textsubscript{2}H\textsubscript{2} abundance by up to $\sim$20\%, the latter two would both act to actually further increase the auroral C\textsubscript{2}H\textsubscript{2} abundance. A similar trend may exist for the northern auroral oval, but the poor spatial coverage and signal-to-noise ratio prevent us from drawing any firm conclusions from the Juno UVS data. 

\section{Discussion}
\label{sec:discussion}

In Section~\ref{sec:results}, we showed that the shape of Jupiter's reflectance spectrum varies significantly across the polar regions of the planet, and that this change in shape can be used to infer spatial variability in the C\textsubscript{2}H\textsubscript{2} abundance. As described in \citeA{giles21b}, the UVS reflected sunlight spectrum is sensitive to the total column abundance down to the 5--50 mbar region, but the C\textsubscript{2}H\textsubscript{2} number density peaks at 0.55 mbar \cite{moses05} and it is therefore this part of the atmosphere that has the strongest influence on the spectral shape. In the south pole, the C\textsubscript{2}H\textsubscript{2} abundance generally decreases towards higher latitudes, with the exception of the auroral region; within the southern auroral oval, the C\textsubscript{2}H\textsubscript{2} abundance is higher than the abundance at low latitudes and is a factor of $\sim$3 higher than adjacent quiescent, non-auroral longitudes. It is possible that there is also an enhancement in the C\textsubscript{6}H\textsubscript{6} abundance within the auroral ovals, as shown from infrared observations by~\citeA{sinclair19}, but the UVS data cannot independently confirm this.

The decrease in the C\textsubscript{2}H\textsubscript{2} abundance towards higher latitudes is consistent with previous studies that have performed full radiative transfer retrievals as a function of latitude. \citeA{giles21b} used zonally-averaged Juno UVS data to study the latitudinal variability of C\textsubscript{2}H\textsubscript{2} and found that the abundance decreased by a factor of 2--4 towards the poles. Similar results were also obtained using infrared observations from the Cassini spacecraft \cite{nixon07} and from ground-based telescopes \cite{fletcher16,melin18}.

The enhancement of C\textsubscript{2}H\textsubscript{2} within the southern polar auroral region is similar to results from \citeA{drossart86} and \citeA{sinclair17b,sinclair18}. \citeA{drossart86} analyzed FPGS data from NASA's Infrared Telescope Facility (IRTF) and found that there was a localized region of strong C\textsubscript{2}H\textsubscript{2} emission that coincided with the location of Jupiter's northern auroral oval. They concluded that this could be due to either a thermal anomaly or an enhancement in C\textsubscript{2}H\textsubscript{2}. More recently, \citeA{sinclair17b,sinclair18} used infrared observations from spacecraft and ground-based telescopes to simultaneously retrieve temperatures and hydrocarbon abundances in Jupiter's polar regions. \citeA{sinclair17b} used data from Voyager IRIS and Cassini CIRS, while \citeA{sinclair18} performed similar analyses using data from the TEXES spectrograph at the IRTF. All three sets of infrared observations showed an enhancement in the C\textsubscript{2}H\textsubscript{2} abundance within the northern auroral oval, relative to the quiescent, non-auroral longitudes, although \citeA{sinclair17b,sinclair18} do note that there is some degeneracy between the thermal profile and the C\textsubscript{2}H\textsubscript{2} abundance.

The results presented in this study build upon the previous works in two ways. Firstly, \citeA{drossart86} and \citeA{sinclair17b,sinclair18} found C\textsubscript{2}H\textsubscript{2} enhancements in the northern polar auroral region, but the southern aurora could not be clearly seen in the observations, because the viewing geometry in each case was close to equatorial and the southern auroral oval is restricted to high latitudes. In contrast, Juno UVS has significantly better coverage at the south pole than the north. Secondly, this study makes use of observations from a completely different spectral region; unlike previous infrared observations, the UV spectra used in this study are not sensitive to the temperature of the atmosphere so there is no degeneracy between the inferred C\textsubscript{2}H\textsubscript{2} abundance and the thermal profile. The striking similarity between the south polar map shown in Figure~\ref{fig:ratio_map}(a) and the north polar maps presented in \citeA{sinclair17b,sinclair18} acts to both confirm the latter's results using an independent spectral region and also show that the same phenomenon exists at both poles.

Stratospheric C\textsubscript{2}H\textsubscript{2} is thought to mainly form on Jupiter from photochemical by-products following methane photolysis \cite <e.g.,>{gladstone96}. The peak C\textsubscript{2}H\textsubscript{2} production regions are $4\times10^{-3}$\,mbar and 0.4\,mbar, and the freshly formed acetylene then diffuses downward to higher pressure levels \cite{hue18b}. The decrease in the C\textsubscript{2}H\textsubscript{2} abundance from the equator to the poles can be explained by the decrease in the solar insolation at high latitudes, and this observed latitudinal trend agrees with the results of photochemical models \cite{gladstone96,hue18b}. However, the low solar insolation rates at Jupiter's poles mean that an alternative mechanism is required to explain the high C\textsubscript{2}H\textsubscript{2} abundances within the auroral regions that were observed in both this work and in the previous infrared studies. The clear correlation between the location of Jupiter's polar aurora, which are located poleward of the main auroral oval, and the region of enhanced C\textsubscript{2}H\textsubscript{2} absorption suggests that the upper-atmosphere chemistry is being modified by the influx of charged auroral particles and the increase in ion-neutral and electron recombination reactions favors the production of C\textsubscript{2}H\textsubscript{2} \cite{sinclair17b}. 

The majority of Jovian photochemical models neglect ion-neutral chemistry and while \citeA{kim94} did study the effect of hydrocarbon ions on the neutral ones, they focus on photoionization from solar extreme ultraviolet photons in the low- and mid-latitudes and make use of chemical networks that have since been updated, including updated reaction rates, additional chemical pathways, and updated branching ratios. In order to explore the exact mechanism by which auroral chemistry may favor C\textsubscript{2}H\textsubscript{2} production, we can instead consider an analogy with Titan's hydrocarbon chemistry, as discussed in \citeA{sinclair17b}. Several ion-neutral photochemical models of Titan's atmosphere have found that electron precipitation triggers the formation of the key ion CH\textsubscript{3}$^+$, which in turn leads to the formation of C\textsubscript{2}H\textsubscript{5}$^+$ \cite{delahaye08, Loison15, Dobrijevic16}. C\textsubscript{2}H\textsubscript{5}$^+$ can then lead to the formation of C\textsubscript{2}H\textsubscript{2} directly, through

\begin{center}
 \item C\textsubscript{2}H\textsubscript{5}$^+$ + e$^-$ 	$\rightarrow$ C\textsubscript{2}H\textsubscript{2} + H\textsubscript{2} + H
\end{center}

\noindent or indirectly, by first forming C\textsubscript{2}H\textsubscript{4}

\begin{center}
 \item C\textsubscript{2}H\textsubscript{5}$^+$ + e$^-$ 	$\rightarrow$ C\textsubscript{2}H\textsubscript{4} + H 
  \item C\textsubscript{2}H\textsubscript{5}$^+$ + HCN 	$\rightarrow$ C\textsubscript{2}H\textsubscript{4} + HCNH$^+$
\end{center}

\noindent which is then photolyzed to produce C\textsubscript{2}H\textsubscript{2} and H\textsubscript{2}. A similar pathway could lead to the enhanced production of C\textsubscript{2}H\textsubscript{2} within Jupiter's auroral regions.

\citeA{kim94} suggested that C\textsubscript{2}H\textsubscript{2} would be easily destroyed in Jupiter's auroral regions due to ion chemistry, due to the reaction

\begin{center}
 CH\textsubscript{5}$^+$ + C\textsubscript{2}H\textsubscript{2} $\rightarrow$ C\textsubscript{2}H\textsubscript{3}$^+$ + CH\textsubscript{4}
\end{center}

In \citeA{Dobrijevic16}, the rate used for this reaction was 5\% slower than the rate used in \citeA{kim94}, leading to a slightly slower C\textsubscript{2}H\textsubscript{2} destruction rate. However, the reaction leading to the formation of CH\textsubscript{3}$^+$, the main building block of C\textsubscript{2}H\textsubscript{5}$^+$, was taken to be 9\% slower in \citeA{Dobrijevic16}, which might also decrease the C\textsubscript{2}H\textsubscript{2} production rate relative to \citeA{kim94}. While enhanced electron precipitation is a plausible driving mechanism for the enhanced C\textsubscript{2}H\textsubscript{2} abundances, the development of an ion-neutral photochemical model of Jupiter's polar regions, using the most recent reaction rates, is needed in order to reach a firm conclusion.

Both this study and \citeA{sinclair17b,sinclair18} have shown that the C\textsubscript{2}H\textsubscript{2} enhancement at Jupiter's poles is longitudinally-restricted to the auroras, instead of being well-mixed across the entire polar region. Jupiter's main auroral regions are encircled by electrojets \cite{rego99}, which have speeds of 1--2 kms\textsuperscript{-1 } and extend from the sub-\textmu bar level \cite{chaufray11} down to at least the middle stratosphere \cite{cavalie21}. One possibility is that these electrojets act as a dynamical barrier to limit the loss of materials produced within the auroras. Alternatively, C\textsubscript{2}H\textsubscript{2} may be efficiently transported outside of the auroral region into the quiescent region, but once in a location where neutral chemistry dominates, it may be rapidly converted into other chemical species such as C\textsubscript{2}H\textsubscript{6} before it can become homogenized across the entire polar region \cite{sinclair17b}. This latter scenario would also explain the increasing C\textsubscript{2}H\textsubscript{6} abundance from the equator to the poles observed by Cassini CIRS \cite{nixon07}, which cannot be explained by a combination of neutral photochemistry and dynamics alone \cite{hue18b}.

In order to obtain sufficient signal-to-noise ratio for this study, we combined observations from the entire Juno mission to date, a period of over 6 years. This means that we are unable to study any temporal variability in the C\textsubscript{2}H\textsubscript{2} auroral enhancement. If the C\textsubscript{2}H\textsubscript{2} auroral enhancement is caused by the influx of charged particles, we might expect the abundance to vary with time, depending on the strength and size of the auroras, which in turn may vary depending on both solar activity and internal factors within the Jovian system~\cite<e.g.>{bonfond12,nichols17,yao22}. The extent to which the C\textsubscript{2}H\textsubscript{2} abundance varies with time may provide additional constraints on the chemical lifetimes and transport rates in the auroral region. Future studies may be able to examine the temporal variability of auroral C\textsubscript{2}H\textsubscript{2}; long-term ground-based infrared observations can be used to track the C\textsubscript{2}H\textsubscript{2} emissions with time~\cite{sinclair18}, and the next-generation UVS instrument on the JUICE mission~\cite<Jupiter Icy Moons Explorer,>{grasset13} is expected to make observations of Jupiter's high latitudes starting in 2031 and will allow a direct comparison with these Juno UVS observations from 2016--2022.

\section{Conclusions}
\label{sec:conclusions}

Reflected sunlight observations from the UVS instrument on the Juno spacecraft show enhanced C\textsubscript{2}H\textsubscript{2} absorption within Jupiter's southern auroral oval. This builds on previous infrared observations which have shown higher C\textsubscript{2}H\textsubscript{2} abundances within the northern auroral oval \cite{sinclair17b,sinclair18}. Our ultraviolet observations both provide a confirmation for these previous results, making use of an independent spectral region, and show that the same phenomenon exists at both poles. Future studies are required to study whether the C\textsubscript{2}H\textsubscript{2} enhancement varies with time.

The localized enhancement of C\textsubscript{2}H\textsubscript{2} is likely caused by the influx of charged particles within Jupiter's auroras, which lead to increased ion-neutral and electron recombination reactions. Ion-neutral photochemical models of Titan's atmosphere provide insight into possible pathways that would favor the production of C\textsubscript{2}H\textsubscript{2} in Jupiter's auroral regions. The development of ion-neutral photochemical models of Jupiter, using the recent chemical reactions rates, is an important next step in understanding the influence of the auroras on Jupiter's polar composition.

\section*{Acknowledgements}

We are grateful to NASA and contributing institutions, which have made the Juno mission possible. This work was funded by NASA's New Frontiers Program for Juno via contract with the Southwest Research Institute. BB is a Research Associate of the Fonds de la Recherche Scientifique - FNRS. 

\section*{Data Availability Statement}

The Juno UVS data used in this paper are archived in NASA's Planetary Data System Atmospheres Node: https://pds-atmospheres.nmsu.edu/PDS/data/jnouvs\_3001 \cite{trantham14}. The data used to produce the figures in this paper are available in \citeA{giles23_data}. The NEMESIS radiative transfer and retrieval tool is available from \citeA{irwin20_nemesis}.


\begin{thebibliography}{}

\bibitem [\protect \citeauthoryear {%
Bolton%
\ \protect \BOthers {.}}{%
Bolton%
\ \protect \BOthers {.}}{%
{\protect \APACyear {2017}}%
}]{%
Bolton17}
\APACinsertmetastar {%
Bolton17}%
\begin{APACrefauthors}%
Bolton, S\BPBI J.%
, Lunine, J.%
, Stevenson, D.%
, Connerney, J\BPBI E\BPBI P.%
, Levin, S.%
, Owen, T\BPBI C.%
\BDBL {}others%
\end{APACrefauthors}%
\unskip\
\newblock
\APACrefYearMonthDay{2017}{}{}.
\newblock
{\BBOQ}\APACrefatitle {{The Juno Mission}} {{The Juno Mission}}.{\BBCQ}
\newblock
\APACjournalVolNumPages{Space Science Reviews}{213}{1-4}{5--37}.
\PrintBackRefs{\CurrentBib}

\bibitem [\protect \citeauthoryear {%
Bonfond%
\ \protect \BOthers {.}}{%
Bonfond%
\ \protect \BOthers {.}}{%
{\protect \APACyear {2012}}%
}]{%
bonfond12}
\APACinsertmetastar {%
bonfond12}%
\begin{APACrefauthors}%
Bonfond, B.%
, Grodent, D.%
, G{\'e}rard, J\BHBI C.%
, Stallard, T.%
, Clarke, J\BPBI T.%
, Yoneda, M.%
\BDBL {}Gustin, J.%
\end{APACrefauthors}%
\unskip\
\newblock
\APACrefYearMonthDay{2012}{}{}.
\newblock
{\BBOQ}\APACrefatitle {Auroral evidence of {Io's} control over the
  magnetosphere of {Jupiter}} {Auroral evidence of {Io's} control over the
  magnetosphere of {Jupiter}}.{\BBCQ}
\newblock
\APACjournalVolNumPages{Geophysical Research Letters}{39}{1}{L01105}.
\PrintBackRefs{\CurrentBib}

\bibitem [\protect \citeauthoryear {%
Bonfond%
\ \protect \BOthers {.}}{%
Bonfond%
\ \protect \BOthers {.}}{%
{\protect \APACyear {2017}}%
}]{%
bonfond17b}
\APACinsertmetastar {%
bonfond17b}%
\begin{APACrefauthors}%
Bonfond, B.%
, Saur, J.%
, Grodent, D.%
, Badman, S\BPBI V.%
, Bisikalo, D.%
, Shematovich, V.%
\BDBL {}Radioti, A.%
\end{APACrefauthors}%
\unskip\
\newblock
\APACrefYearMonthDay{2017}{}{}.
\newblock
{\BBOQ}\APACrefatitle {The tails of the satellite auroral footprints at
  {Jupiter}} {The tails of the satellite auroral footprints at
  {Jupiter}}.{\BBCQ}
\newblock
\APACjournalVolNumPages{Journal of Geophysical Research: Space
  Physics}{122}{8}{7985--7996}.
\PrintBackRefs{\CurrentBib}

\bibitem [\protect \citeauthoryear {%
Caldwell%
, Tokunaga%
\BCBL {}\ \BBA {} Gillett%
}{%
Caldwell%
\ \protect \BOthers {.}}{%
{\protect \APACyear {1980}}%
}]{%
caldwell80}
\APACinsertmetastar {%
caldwell80}%
\begin{APACrefauthors}%
Caldwell, J.%
, Tokunaga, A\BPBI T.%
\BCBL {}\ \BBA {} Gillett, F\BPBI C.%
\end{APACrefauthors}%
\unskip\
\newblock
\APACrefYearMonthDay{1980}{}{}.
\newblock
{\BBOQ}\APACrefatitle {Possible infrared aurorae on {Jupiter}} {Possible
  infrared aurorae on {Jupiter}}.{\BBCQ}
\newblock
\APACjournalVolNumPages{Icarus}{44}{3}{667--675}.
\PrintBackRefs{\CurrentBib}

\bibitem [\protect \citeauthoryear {%
Cavali{\'e}%
\ \protect \BOthers {.}}{%
Cavali{\'e}%
\ \protect \BOthers {.}}{%
{\protect \APACyear {2021}}%
}]{%
cavalie21}
\APACinsertmetastar {%
cavalie21}%
\begin{APACrefauthors}%
Cavali{\'e}, T.%
, Benmahi, B.%
, Hue, V.%
, Moreno, R.%
, Lellouch, E.%
, Fouchet, T.%
\BDBL {}others%
\end{APACrefauthors}%
\unskip\
\newblock
\APACrefYearMonthDay{2021}{}{}.
\newblock
{\BBOQ}\APACrefatitle {First direct measurement of auroral and equatorial jets
  in the stratosphere of {Jupiter}} {First direct measurement of auroral and
  equatorial jets in the stratosphere of {Jupiter}}.{\BBCQ}
\newblock
\APACjournalVolNumPages{Astronomy \& Astrophysics}{647}{}{L8}.
\PrintBackRefs{\CurrentBib}

\bibitem [\protect \citeauthoryear {%
Chaufray%
\ \protect \BOthers {.}}{%
Chaufray%
\ \protect \BOthers {.}}{%
{\protect \APACyear {2011}}%
}]{%
chaufray11}
\APACinsertmetastar {%
chaufray11}%
\begin{APACrefauthors}%
Chaufray, J\BHBI Y.%
, Greathouse, T\BPBI K.%
, Gladstone, G\BPBI R.%
, Waite, J\BPBI H.%
, Maillard, J\BHBI P.%
, Majeed, T.%
\BDBL {}Drossart, P.%
\end{APACrefauthors}%
\unskip\
\newblock
\APACrefYearMonthDay{2011}{}{}.
\newblock
{\BBOQ}\APACrefatitle {Spectro-imaging observations of {Jupiter's} 2 $\mu$m
  auroral emission. {II}: Thermospheric winds} {Spectro-imaging observations of
  {Jupiter's} 2 $\mu$m auroral emission. {II}: Thermospheric winds}.{\BBCQ}
\newblock
\APACjournalVolNumPages{Icarus}{211}{2}{1233--1241}.
\PrintBackRefs{\CurrentBib}

\bibitem [\protect \citeauthoryear {%
De~La~Haye%
, Waite~Jr%
, Cravens%
, Robertson%
\BCBL {}\ \BBA {} Lebonnois%
}{%
De~La~Haye%
\ \protect \BOthers {.}}{%
{\protect \APACyear {2008}}%
}]{%
delahaye08}
\APACinsertmetastar {%
delahaye08}%
\begin{APACrefauthors}%
De~La~Haye, V.%
, Waite~Jr, J\BPBI H.%
, Cravens, T\BPBI E.%
, Robertson, I\BPBI P.%
\BCBL {}\ \BBA {} Lebonnois, S.%
\end{APACrefauthors}%
\unskip\
\newblock
\APACrefYearMonthDay{2008}{}{}.
\newblock
{\BBOQ}\APACrefatitle {Coupled ion and neutral rotating model of {Titan's}
  upper atmosphere} {Coupled ion and neutral rotating model of {Titan's} upper
  atmosphere}.{\BBCQ}
\newblock
\APACjournalVolNumPages{Icarus}{197}{1}{110--136}.
\PrintBackRefs{\CurrentBib}

\bibitem [\protect \citeauthoryear {%
Dobrijevic%
, Loison%
, Hickson%
\BCBL {}\ \BBA {} Gronoff%
}{%
Dobrijevic%
\ \protect \BOthers {.}}{%
{\protect \APACyear {2016}}%
}]{%
Dobrijevic16}
\APACinsertmetastar {%
Dobrijevic16}%
\begin{APACrefauthors}%
Dobrijevic, M.%
, Loison, J\BPBI C.%
, Hickson, K\BPBI M.%
\BCBL {}\ \BBA {} Gronoff, G.%
\end{APACrefauthors}%
\unskip\
\newblock
\APACrefYearMonthDay{2016}{}{}.
\newblock
{\BBOQ}\APACrefatitle {{1D}-coupled photochemical model of neutrals, cations
  and anions in the atmosphere of {Titan}} {{1D}-coupled photochemical model of
  neutrals, cations and anions in the atmosphere of {Titan}}.{\BBCQ}
\newblock
\APACjournalVolNumPages{Icarus}{268}{}{313--339}.
\PrintBackRefs{\CurrentBib}

\bibitem [\protect \citeauthoryear {%
Drossart%
\ \protect \BOthers {.}}{%
Drossart%
\ \protect \BOthers {.}}{%
{\protect \APACyear {1993}}%
}]{%
drossart93b}
\APACinsertmetastar {%
drossart93b}%
\begin{APACrefauthors}%
Drossart, P.%
, B{\'e}zard, B.%
, Atreya, S\BPBI K.%
, Bishop, J.%
, Waite~Jr, J.%
\BCBL {}\ \BBA {} Boice, D.%
\end{APACrefauthors}%
\unskip\
\newblock
\APACrefYearMonthDay{1993}{}{}.
\newblock
{\BBOQ}\APACrefatitle {Thermal profiles in the auroral regions of {Jupiter}}
  {Thermal profiles in the auroral regions of {Jupiter}}.{\BBCQ}
\newblock
\APACjournalVolNumPages{Journal of Geophysical Research:
  Planets}{98}{E10}{18803--18811}.
\PrintBackRefs{\CurrentBib}

\bibitem [\protect \citeauthoryear {%
Drossart%
\ \protect \BOthers {.}}{%
Drossart%
\ \protect \BOthers {.}}{%
{\protect \APACyear {1986}}%
}]{%
drossart86}
\APACinsertmetastar {%
drossart86}%
\begin{APACrefauthors}%
Drossart, P.%
, B{\'e}zard, B.%
, Atreya, S\BPBI K.%
, Lacy, J.%
, Serabyn, E.%
, Tokunaga, A.%
\BCBL {}\ \BBA {} Encrenaz, T.%
\end{APACrefauthors}%
\unskip\
\newblock
\APACrefYearMonthDay{1986}{}{}.
\newblock
{\BBOQ}\APACrefatitle {Enhanced acetylene emission near the north pole of
  {Jupiter}} {Enhanced acetylene emission near the north pole of
  {Jupiter}}.{\BBCQ}
\newblock
\APACjournalVolNumPages{Icarus}{66}{3}{610--618}.
\PrintBackRefs{\CurrentBib}

\bibitem [\protect \citeauthoryear {%
Fletcher%
\ \protect \BOthers {.}}{%
Fletcher%
\ \protect \BOthers {.}}{%
{\protect \APACyear {2016}}%
}]{%
fletcher16}
\APACinsertmetastar {%
fletcher16}%
\begin{APACrefauthors}%
Fletcher, L\BPBI N.%
, Greathouse, T\BPBI K.%
, Orton, G\BPBI S.%
, Sinclair, J\BPBI A.%
, Giles, R\BPBI S.%
, Irwin, P\BPBI G\BPBI J.%
\BCBL {}\ \BBA {} Encrenaz, T.%
\end{APACrefauthors}%
\unskip\
\newblock
\APACrefYearMonthDay{2016}{}{}.
\newblock
{\BBOQ}\APACrefatitle {Mid-infrared mapping of {Jupiter's} temperatures,
  aerosol opacity and chemical distributions with {IRTF/TEXES}} {Mid-infrared
  mapping of {Jupiter's} temperatures, aerosol opacity and chemical
  distributions with {IRTF/TEXES}}.{\BBCQ}
\newblock
\APACjournalVolNumPages{Icarus}{278}{}{128--161}.
\PrintBackRefs{\CurrentBib}

\bibitem [\protect \citeauthoryear {%
Giles%
}{%
Giles%
}{%
{\protect \APACyear {2023}}%
}]{%
giles23_data}
\APACinsertmetastar {%
giles23_data}%
\begin{APACrefauthors}%
Giles, R\BPBI S.%
\end{APACrefauthors}%
\unskip\
\newblock
\APACrefYearMonthDay{2023}{}{}.
\newblock
\APACrefbtitle {Enhanced {C\textsubscript{2}H\textsubscript{2}} absorption
  within {Jupiter's} southern auroral oval from {Juno UVS} observations.}
  {Enhanced {C\textsubscript{2}H\textsubscript{2}} absorption within
  {Jupiter's} southern auroral oval from {Juno UVS} observations.}
\newblock
\APAChowpublished {Mendeley Data, V1}.
\newblock
\begin{APACrefDOI} \doi{10.17632/ttyx24ykcx.1} \end{APACrefDOI}
\PrintBackRefs{\CurrentBib}

\bibitem [\protect \citeauthoryear {%
Giles%
\ \protect \BOthers {.}}{%
Giles%
\ \protect \BOthers {.}}{%
{\protect \APACyear {2021}}%
}]{%
giles21b}
\APACinsertmetastar {%
giles21b}%
\begin{APACrefauthors}%
Giles, R\BPBI S.%
, Greathouse, T\BPBI K.%
, Hue, V.%
, Gladstone, G\BPBI R.%
, Melin, H.%
, Fletcher, L\BPBI N.%
\BDBL {}Levin, S\BPBI M.%
\end{APACrefauthors}%
\unskip\
\newblock
\APACrefYearMonthDay{2021}{}{}.
\newblock
{\BBOQ}\APACrefatitle {Meridional variations of
  {C\textsubscript{2}H\textsubscript{2}} in {Jupiter's} stratosphere from {Juno
  UVS} observations} {Meridional variations of
  {C\textsubscript{2}H\textsubscript{2}} in {Jupiter's} stratosphere from {Juno
  UVS} observations}.{\BBCQ}
\newblock
\APACjournalVolNumPages{Journal of Geophysical Research:
  Planets}{126}{8}{e2021JE006928}.
\PrintBackRefs{\CurrentBib}

\bibitem [\protect \citeauthoryear {%
Gladstone%
, Allen%
\BCBL {}\ \BBA {} Yung%
}{%
Gladstone%
\ \protect \BOthers {.}}{%
{\protect \APACyear {1996}}%
}]{%
gladstone96}
\APACinsertmetastar {%
gladstone96}%
\begin{APACrefauthors}%
Gladstone, G\BPBI R.%
, Allen, M.%
\BCBL {}\ \BBA {} Yung, Y\BPBI L.%
\end{APACrefauthors}%
\unskip\
\newblock
\APACrefYearMonthDay{1996}{}{}.
\newblock
{\BBOQ}\APACrefatitle {Hydrocarbon photochemistry in the upper atmosphere of
  {Jupiter}} {Hydrocarbon photochemistry in the upper atmosphere of
  {Jupiter}}.{\BBCQ}
\newblock
\APACjournalVolNumPages{Icarus}{119}{1}{1--52}.
\PrintBackRefs{\CurrentBib}

\bibitem [\protect \citeauthoryear {%
Gladstone%
\ \protect \BOthers {.}}{%
Gladstone%
\ \protect \BOthers {.}}{%
{\protect \APACyear {2019}}%
}]{%
gladstone19}
\APACinsertmetastar {%
gladstone19}%
\begin{APACrefauthors}%
Gladstone, G\BPBI R.%
, Greathouse, T\BPBI K.%
, Versteeg, M\BPBI H.%
, Hue, V.%
, Kammer, J\BPBI A.%
, Davis, M\BPBI W.%
\BDBL {}J, C.%
\end{APACrefauthors}%
\unskip\
\newblock
\APACrefYearMonthDay{2019}{{\APACmonth{10}}}{}.
\newblock
{\BBOQ}\APACrefatitle {Recent {Juno-UVS} Observations of {Jupiter's} Auroras}
  {Recent {Juno-UVS} observations of {Jupiter's} auroras}.{\BBCQ}
\newblock
\BIn{} \APACrefbtitle {{EPSC/DPS Joint Meeting Abstracts}} {{EPSC/DPS Joint
  Meeting Abstracts}}\ (\BVOL~13).
\PrintBackRefs{\CurrentBib}

\bibitem [\protect \citeauthoryear {%
Gladstone%
\ \protect \BOthers {.}}{%
Gladstone%
\ \protect \BOthers {.}}{%
{\protect \APACyear {2017}}%
}]{%
gladstone17}
\APACinsertmetastar {%
gladstone17}%
\begin{APACrefauthors}%
Gladstone, G\BPBI R.%
, Persyn, S\BPBI C.%
, Eterno, J\BPBI S.%
, Walther, B\BPBI C.%
, Slater, D\BPBI C.%
, Davis, M\BPBI W.%
\BDBL {}others%
\end{APACrefauthors}%
\unskip\
\newblock
\APACrefYearMonthDay{2017}{}{}.
\newblock
{\BBOQ}\APACrefatitle {The ultraviolet spectrograph on {NASA's Juno} mission}
  {The ultraviolet spectrograph on {NASA's Juno} mission}.{\BBCQ}
\newblock
\APACjournalVolNumPages{Space Science Reviews}{213}{1-4}{447--473}.
\PrintBackRefs{\CurrentBib}

\bibitem [\protect \citeauthoryear {%
Grasset%
\ \protect \BOthers {.}}{%
Grasset%
\ \protect \BOthers {.}}{%
{\protect \APACyear {2013}}%
}]{%
grasset13}
\APACinsertmetastar {%
grasset13}%
\begin{APACrefauthors}%
Grasset, O.%
, Dougherty, M\BPBI K.%
, Coustenis, A.%
, Bunce, E\BPBI J.%
, Erd, C.%
, Titov, D.%
\BDBL {}others%
\end{APACrefauthors}%
\unskip\
\newblock
\APACrefYearMonthDay{2013}{}{}.
\newblock
{\BBOQ}\APACrefatitle {{JUpiter ICy moons Explorer (JUICE)}: An {ESA} mission
  to orbit {Ganymede} and to characterise the {Jupiter} system} {{JUpiter ICy
  moons Explorer (JUICE)}: An {ESA} mission to orbit {Ganymede} and to
  characterise the {Jupiter} system}.{\BBCQ}
\newblock
\APACjournalVolNumPages{Planetary and Space Science}{78}{}{1--21}.
\PrintBackRefs{\CurrentBib}

\bibitem [\protect \citeauthoryear {%
Greathouse%
\ \protect \BOthers {.}}{%
Greathouse%
\ \protect \BOthers {.}}{%
{\protect \APACyear {2013}}%
}]{%
greathouse13}
\APACinsertmetastar {%
greathouse13}%
\begin{APACrefauthors}%
Greathouse, T\BPBI K.%
, Gladstone, G\BPBI R.%
, Davis, M\BPBI W.%
, Slater, D\BPBI C.%
, Versteeg, M\BPBI H.%
, Persson, K\BPBI B.%
\BDBL {}Eterno, J\BPBI S.%
\end{APACrefauthors}%
\unskip\
\newblock
\APACrefYearMonthDay{2013}{}{}.
\newblock
{\BBOQ}\APACrefatitle {Performance results from in-flight commissioning of the
  {Juno Ultraviolet Spectrograph (Juno-UVS)}} {Performance results from
  in-flight commissioning of the {Juno Ultraviolet Spectrograph
  (Juno-UVS)}}.{\BBCQ}
\newblock
\BIn{} \APACrefbtitle {{UV, X-Ray, and Gamma-Ray Space Instrumentation for
  Astronomy XVIII}} {{UV, X-Ray, and Gamma-Ray Space Instrumentation for
  Astronomy XVIII}}\ (\BVOL\ 8859, \BPG~88590T).
\PrintBackRefs{\CurrentBib}

\bibitem [\protect \citeauthoryear {%
Grodent%
}{%
Grodent%
}{%
{\protect \APACyear {2015}}%
}]{%
grodent15}
\APACinsertmetastar {%
grodent15}%
\begin{APACrefauthors}%
Grodent, D.%
\end{APACrefauthors}%
\unskip\
\newblock
\APACrefYearMonthDay{2015}{}{}.
\newblock
{\BBOQ}\APACrefatitle {A brief review of ultraviolet auroral emissions on giant
  planets} {A brief review of ultraviolet auroral emissions on giant
  planets}.{\BBCQ}
\newblock
\APACjournalVolNumPages{Space Science Reviews}{187}{1}{23--50}.
\PrintBackRefs{\CurrentBib}

\bibitem [\protect \citeauthoryear {%
Hue%
\ \protect \BOthers {.}}{%
Hue%
\ \protect \BOthers {.}}{%
{\protect \APACyear {2021}}%
}]{%
hue21b}
\APACinsertmetastar {%
hue21b}%
\begin{APACrefauthors}%
Hue, V.%
, Giles, R\BPBI S.%
, Gladstone, G\BPBI R.%
, Greathouse, T\BPBI K.%
, Davis, M\BPBI W.%
, Kammer, J\BPBI A.%
\BCBL {}\ \BBA {} Versteeg, M\BPBI H.%
\end{APACrefauthors}%
\unskip\
\newblock
\APACrefYearMonthDay{2021}{}{}.
\newblock
{\BBOQ}\APACrefatitle {Updated radiometric and wavelength calibration of the
  {Juno} ultraviolet spectrograph} {Updated radiometric and wavelength
  calibration of the {Juno} ultraviolet spectrograph}.{\BBCQ}
\newblock
\APACjournalVolNumPages{Journal of Astronomical Telescopes, Instruments, and
  Systems}{7}{4}{044003}.
\PrintBackRefs{\CurrentBib}

\bibitem [\protect \citeauthoryear {%
Hue%
\ \protect \BOthers {.}}{%
Hue%
\ \protect \BOthers {.}}{%
{\protect \APACyear {2019}}%
}]{%
hue19b}
\APACinsertmetastar {%
hue19b}%
\begin{APACrefauthors}%
Hue, V.%
, Gladstone, G\BPBI R.%
, Greathouse, T\BPBI K.%
, Kammer, J\BPBI A.%
, Davis, M\BPBI W.%
, Bonfond, B.%
\BDBL {}others%
\end{APACrefauthors}%
\unskip\
\newblock
\APACrefYearMonthDay{2019}{}{}.
\newblock
{\BBOQ}\APACrefatitle {In-flight characterization and calibration of the
  {Juno}-ultraviolet spectrograph {(Juno-UVS)}} {In-flight characterization and
  calibration of the {Juno}-ultraviolet spectrograph {(Juno-UVS)}}.{\BBCQ}
\newblock
\APACjournalVolNumPages{The Astronomical Journal}{157}{2}{90}.
\PrintBackRefs{\CurrentBib}

\bibitem [\protect \citeauthoryear {%
Hue%
, Hersant%
, Cavali{\'e}%
, Dobrijevic%
\BCBL {}\ \BBA {} Sinclair%
}{%
Hue%
\ \protect \BOthers {.}}{%
{\protect \APACyear {2018}}%
}]{%
hue18b}
\APACinsertmetastar {%
hue18b}%
\begin{APACrefauthors}%
Hue, V.%
, Hersant, F.%
, Cavali{\'e}, T.%
, Dobrijevic, M.%
\BCBL {}\ \BBA {} Sinclair, J.%
\end{APACrefauthors}%
\unskip\
\newblock
\APACrefYearMonthDay{2018}{}{}.
\newblock
{\BBOQ}\APACrefatitle {Photochemistry, mixing and transport in {Jupiter's}
  stratosphere constrained by {Cassini}} {Photochemistry, mixing and transport
  in {Jupiter's} stratosphere constrained by {Cassini}}.{\BBCQ}
\newblock
\APACjournalVolNumPages{Icarus}{307}{}{106--123}.
\PrintBackRefs{\CurrentBib}

\bibitem [\protect \citeauthoryear {%
Irwin%
}{%
Irwin%
}{%
{\protect \APACyear {2020}}%
}]{%
irwin20_nemesis}
\APACinsertmetastar {%
irwin20_nemesis}%
\begin{APACrefauthors}%
Irwin, P\BPBI G\BPBI J.%
\end{APACrefauthors}%
\unskip\
\newblock
\APACrefYearMonthDay{2020}{}{}.
\newblock
\APACrefbtitle {{NEMESIS/Radtrancode Software (Version 1.0)}.}
  {{NEMESIS/Radtrancode Software (Version 1.0)}.}
\newblock
\APAChowpublished {Zenodo}.
\newblock
\begin{APACrefDOI} \doi{10.5281/zenodo.4303976} \end{APACrefDOI}
\PrintBackRefs{\CurrentBib}

\bibitem [\protect \citeauthoryear {%
Irwin%
\ \protect \BOthers {.}}{%
Irwin%
\ \protect \BOthers {.}}{%
{\protect \APACyear {2008}}%
}]{%
irwin08}
\APACinsertmetastar {%
irwin08}%
\begin{APACrefauthors}%
Irwin, P\BPBI G\BPBI J.%
, Teanby, N\BPBI A.%
, de Kok, R.%
, Fletcher, L\BPBI N.%
, Howett, C\BPBI J\BPBI A.%
, Tsang, C\BPBI C\BPBI C.%
\BDBL {}Parrish, P\BPBI D.%
\end{APACrefauthors}%
\unskip\
\newblock
\APACrefYearMonthDay{2008}{}{}.
\newblock
{\BBOQ}\APACrefatitle {{The NEMESIS planetary atmosphere radiative transfer and
  retrieval tool}} {{The NEMESIS planetary atmosphere radiative transfer and
  retrieval tool}}.{\BBCQ}
\newblock
\APACjournalVolNumPages{Journal of Quantitative Spectroscopy and Radiative
  Transfer}{109}{6}{1136--1150}.
\PrintBackRefs{\CurrentBib}

\bibitem [\protect \citeauthoryear {%
Kammer%
\ \protect \BOthers {.}}{%
Kammer%
\ \protect \BOthers {.}}{%
{\protect \APACyear {2019}}%
}]{%
kammer19}
\APACinsertmetastar {%
kammer19}%
\begin{APACrefauthors}%
Kammer, J\BPBI A.%
, Hue, V.%
, Greathouse, T\BPBI K.%
, Gladstone, G\BPBI R.%
, Davis, M\BPBI W.%
\BCBL {}\ \BBA {} Versteeg, M\BPBI H.%
\end{APACrefauthors}%
\unskip\
\newblock
\APACrefYearMonthDay{2019}{}{}.
\newblock
{\BBOQ}\APACrefatitle {Planning operations in {Jupiter's} high-radiation
  environment: optimization strategies from {Juno}-ultraviolet spectrograph}
  {Planning operations in {Jupiter's} high-radiation environment: optimization
  strategies from {Juno}-ultraviolet spectrograph}.{\BBCQ}
\newblock
\APACjournalVolNumPages{Journal of Astronomical Telescopes, Instruments, and
  Systems}{5}{2}{027001}.
\PrintBackRefs{\CurrentBib}

\bibitem [\protect \citeauthoryear {%
Keller-Rudek%
, Moortgat%
, Sander%
\BCBL {}\ \BBA {} S{\"o}rensen%
}{%
Keller-Rudek%
\ \protect \BOthers {.}}{%
{\protect \APACyear {2013}}%
}]{%
keller-rudek13}
\APACinsertmetastar {%
keller-rudek13}%
\begin{APACrefauthors}%
Keller-Rudek, H.%
, Moortgat, G\BPBI K.%
, Sander, R.%
\BCBL {}\ \BBA {} S{\"o}rensen, R.%
\end{APACrefauthors}%
\unskip\
\newblock
\APACrefYearMonthDay{2013}{}{}.
\newblock
{\BBOQ}\APACrefatitle {The {MPI-Mainz UV/VIS} spectral atlas of gaseous
  molecules of atmospheric interest} {The {MPI-Mainz UV/VIS} spectral atlas of
  gaseous molecules of atmospheric interest}.{\BBCQ}
\newblock
\APACjournalVolNumPages{Earth System Science Data}{5}{2}{365--373}.
\PrintBackRefs{\CurrentBib}

\bibitem [\protect \citeauthoryear {%
S\BPBI J.~Kim%
, Caldwell%
, Rivolo%
, Wagener%
\BCBL {}\ \BBA {} Orton%
}{%
S\BPBI J.~Kim%
\ \protect \BOthers {.}}{%
{\protect \APACyear {1985}}%
}]{%
kim85}
\APACinsertmetastar {%
kim85}%
\begin{APACrefauthors}%
Kim, S\BPBI J.%
, Caldwell, J.%
, Rivolo, A\BPBI R.%
, Wagener, R.%
\BCBL {}\ \BBA {} Orton, G\BPBI S.%
\end{APACrefauthors}%
\unskip\
\newblock
\APACrefYearMonthDay{1985}{}{}.
\newblock
{\BBOQ}\APACrefatitle {Infrared polar brightening on {Jupiter: III.
  Spectrometry} from the {Voyager 1 IRIS} experiment} {Infrared polar
  brightening on {Jupiter: III. Spectrometry} from the {Voyager 1 IRIS}
  experiment}.{\BBCQ}
\newblock
\APACjournalVolNumPages{Icarus}{64}{2}{233--248}.
\PrintBackRefs{\CurrentBib}

\bibitem [\protect \citeauthoryear {%
Y\BPBI H.~Kim%
\ \BBA {} Fox%
}{%
Y\BPBI H.~Kim%
\ \BBA {} Fox%
}{%
{\protect \APACyear {1994}}%
}]{%
kim94}
\APACinsertmetastar {%
kim94}%
\begin{APACrefauthors}%
Kim, Y\BPBI H.%
\BCBT {}\ \BBA {} Fox, J\BPBI L.%
\end{APACrefauthors}%
\unskip\
\newblock
\APACrefYearMonthDay{1994}{}{}.
\newblock
{\BBOQ}\APACrefatitle {The chemistry of hydrocarbon ions in the {Jovian}
  ionosphere} {The chemistry of hydrocarbon ions in the {Jovian}
  ionosphere}.{\BBCQ}
\newblock
\APACjournalVolNumPages{Icarus}{112}{2}{310--325}.
\PrintBackRefs{\CurrentBib}

\bibitem [\protect \citeauthoryear {%
Loison%
\ \protect \BOthers {.}}{%
Loison%
\ \protect \BOthers {.}}{%
{\protect \APACyear {2015}}%
}]{%
Loison15}
\APACinsertmetastar {%
Loison15}%
\begin{APACrefauthors}%
Loison, J\BPBI C.%
, H{\'e}brard, E.%
, Dobrijevic, M.%
, Hickson, K\BPBI M.%
, Caralp, F.%
, Hue, V.%
\BDBL {}B{\'e}nilan, Y.%
\end{APACrefauthors}%
\unskip\
\newblock
\APACrefYearMonthDay{2015}{}{}.
\newblock
{\BBOQ}\APACrefatitle {The neutral photochemistry of nitriles, amines and
  imines in the atmosphere of {Titan}} {The neutral photochemistry of nitriles,
  amines and imines in the atmosphere of {Titan}}.{\BBCQ}
\newblock
\APACjournalVolNumPages{Icarus}{247}{}{218--247}.
\PrintBackRefs{\CurrentBib}

\bibitem [\protect \citeauthoryear {%
Melin%
\ \protect \BOthers {.}}{%
Melin%
\ \protect \BOthers {.}}{%
{\protect \APACyear {2018}}%
}]{%
melin18}
\APACinsertmetastar {%
melin18}%
\begin{APACrefauthors}%
Melin, H.%
, Fletcher, L\BPBI N.%
, Donnelly, P\BPBI T.%
, Greathouse, T\BPBI K.%
, Lacy, J\BPBI H.%
, Orton, G\BPBI S.%
\BDBL {}Irwin, P\BPBI G\BPBI J.%
\end{APACrefauthors}%
\unskip\
\newblock
\APACrefYearMonthDay{2018}{}{}.
\newblock
{\BBOQ}\APACrefatitle {Assessing the long-term variability of acetylene and
  ethane in the stratosphere of {Jupiter}} {Assessing the long-term variability
  of acetylene and ethane in the stratosphere of {Jupiter}}.{\BBCQ}
\newblock
\APACjournalVolNumPages{Icarus}{305}{1}{301--313}.
\PrintBackRefs{\CurrentBib}

\bibitem [\protect \citeauthoryear {%
Melin%
, Fletcher%
, Irwin%
\BCBL {}\ \BBA {} Edgington%
}{%
Melin%
\ \protect \BOthers {.}}{%
{\protect \APACyear {2020}}%
}]{%
melin20}
\APACinsertmetastar {%
melin20}%
\begin{APACrefauthors}%
Melin, H.%
, Fletcher, L\BPBI N.%
, Irwin, P\BPBI G\BPBI J.%
\BCBL {}\ \BBA {} Edgington, S\BPBI G.%
\end{APACrefauthors}%
\unskip\
\newblock
\APACrefYearMonthDay{2020}{}{}.
\newblock
{\BBOQ}\APACrefatitle {{Jupiter} in the Ultraviolet: Acetylene and Ethane
  Abundances in the Stratosphere of {Jupiter} from {Cassini} Observations
  between 0.15 and 0.19 \textmu m} {{Jupiter} in the ultraviolet: Acetylene and
  ethane abundances in the stratosphere of {Jupiter} from {Cassini}
  observations between 0.15 and 0.19 \textmu m}.{\BBCQ}
\newblock
\APACjournalVolNumPages{The Astronomical Journal}{159}{6}{291}.
\PrintBackRefs{\CurrentBib}

\bibitem [\protect \citeauthoryear {%
Moses%
\ \protect \BOthers {.}}{%
Moses%
\ \protect \BOthers {.}}{%
{\protect \APACyear {2005}}%
}]{%
moses05}
\APACinsertmetastar {%
moses05}%
\begin{APACrefauthors}%
Moses, J\BPBI I.%
, Fouchet, T.%
, B{\'e}zard, B.%
, Gladstone, G\BPBI R.%
, Lellouch, E.%
\BCBL {}\ \BBA {} Feuchtgruber, H.%
\end{APACrefauthors}%
\unskip\
\newblock
\APACrefYearMonthDay{2005}{}{}.
\newblock
{\BBOQ}\APACrefatitle {Photochemistry and diffusion in {Jupiter's}
  stratosphere: constraints from {ISO} observations and comparisons with other
  giant planets} {Photochemistry and diffusion in {Jupiter's} stratosphere:
  constraints from {ISO} observations and comparisons with other giant
  planets}.{\BBCQ}
\newblock
\APACjournalVolNumPages{Journal of Geophysical Research:
  Planets}{110}{E8}{E08001}.
\PrintBackRefs{\CurrentBib}

\bibitem [\protect \citeauthoryear {%
Nichols%
\ \protect \BOthers {.}}{%
Nichols%
\ \protect \BOthers {.}}{%
{\protect \APACyear {2017}}%
}]{%
nichols17}
\APACinsertmetastar {%
nichols17}%
\begin{APACrefauthors}%
Nichols, J\BPBI D.%
, Badman, S\BPBI V.%
, Bagenal, F.%
, Bolton, S\BPBI J.%
, Bonfond, B.%
, Bunce, E\BPBI J.%
\BDBL {}others%
\end{APACrefauthors}%
\unskip\
\newblock
\APACrefYearMonthDay{2017}{}{}.
\newblock
{\BBOQ}\APACrefatitle {Response of {Jupiter's} auroras to conditions in the
  interplanetary medium as measured by the {Hubble Space Telescope and Juno}}
  {Response of {Jupiter's} auroras to conditions in the interplanetary medium
  as measured by the {Hubble Space Telescope and Juno}}.{\BBCQ}
\newblock
\APACjournalVolNumPages{Geophysical Research Letters}{44}{15}{7643--7652}.
\PrintBackRefs{\CurrentBib}

\bibitem [\protect \citeauthoryear {%
Nixon%
\ \protect \BOthers {.}}{%
Nixon%
\ \protect \BOthers {.}}{%
{\protect \APACyear {2007}}%
}]{%
nixon07}
\APACinsertmetastar {%
nixon07}%
\begin{APACrefauthors}%
Nixon, C\BPBI A.%
, Achterberg, R\BPBI K.%
, Conrath, B\BPBI J.%
, Irwin, P\BPBI G\BPBI J.%
, Teanby, N\BPBI A.%
, Fouchet, T.%
\BDBL {}others%
\end{APACrefauthors}%
\unskip\
\newblock
\APACrefYearMonthDay{2007}{}{}.
\newblock
{\BBOQ}\APACrefatitle {Meridional variations of
  {C\textsubscript{2}H\textsubscript{2} and
  C\textsubscript{2}H\textsubscript{6}} in {Jupiter's} atmosphere from {Cassini
  CIRS} infrared spectra} {Meridional variations of
  {C\textsubscript{2}H\textsubscript{2} and
  C\textsubscript{2}H\textsubscript{6}} in {Jupiter's} atmosphere from {Cassini
  CIRS} infrared spectra}.{\BBCQ}
\newblock
\APACjournalVolNumPages{Icarus}{188}{1}{47--71}.
\PrintBackRefs{\CurrentBib}

\bibitem [\protect \citeauthoryear {%
O'Donoghue%
\ \protect \BOthers {.}}{%
O'Donoghue%
\ \protect \BOthers {.}}{%
{\protect \APACyear {2021}}%
}]{%
odonoghue21}
\APACinsertmetastar {%
odonoghue21}%
\begin{APACrefauthors}%
O'Donoghue, J.%
, Moore, L.%
, Bhakyapaibul, T.%
, Melin, H.%
, Stallard, T.%
, Connerney, J.%
\BCBL {}\ \BBA {} Tao, C.%
\end{APACrefauthors}%
\unskip\
\newblock
\APACrefYearMonthDay{2021}{}{}.
\newblock
{\BBOQ}\APACrefatitle {Global upper-atmospheric heating on {Jupiter} by the
  polar aurorae} {Global upper-atmospheric heating on {Jupiter} by the polar
  aurorae}.{\BBCQ}
\newblock
\APACjournalVolNumPages{Nature}{596}{7870}{54--57}.
\PrintBackRefs{\CurrentBib}

\bibitem [\protect \citeauthoryear {%
Rego%
\ \protect \BOthers {.}}{%
Rego%
\ \protect \BOthers {.}}{%
{\protect \APACyear {1999}}%
}]{%
rego99}
\APACinsertmetastar {%
rego99}%
\begin{APACrefauthors}%
Rego, D.%
, Achilleos, N.%
, Stallard, T.%
, Miller, S.%
, Prang{\'e}, R.%
, Dougherty, M.%
\BCBL {}\ \BBA {} Joseph, R\BPBI D.%
\end{APACrefauthors}%
\unskip\
\newblock
\APACrefYearMonthDay{1999}{}{}.
\newblock
{\BBOQ}\APACrefatitle {Supersonic winds in {Jupiter's} aurorae} {Supersonic
  winds in {Jupiter's} aurorae}.{\BBCQ}
\newblock
\APACjournalVolNumPages{Nature}{399}{6732}{121--124}.
\PrintBackRefs{\CurrentBib}

\bibitem [\protect \citeauthoryear {%
Sinclair%
\ \protect \BOthers {.}}{%
Sinclair%
\ \protect \BOthers {.}}{%
{\protect \APACyear {2019}}%
}]{%
sinclair19}
\APACinsertmetastar {%
sinclair19}%
\begin{APACrefauthors}%
Sinclair, J\BPBI A.%
, Moses, J\BPBI I.%
, Hue, V.%
, Greathouse, T\BPBI K.%
, Orton, G\BPBI S.%
, Fletcher, L\BPBI N.%
\BCBL {}\ \BBA {} Irwin, P\BPBI G\BPBI J.%
\end{APACrefauthors}%
\unskip\
\newblock
\APACrefYearMonthDay{2019}{}{}.
\newblock
{\BBOQ}\APACrefatitle {Jupiter's auroral-related stratospheric heating and
  chemistry {III}: Abundances of {C\textsubscript{2}H\textsubscript{4},
  CH\textsubscript{3}C\textsubscript{2}H, C\textsubscript{4}H\textsubscript{2}
  and C\textsubscript{6}H\textsubscript{6} from Voyager-IRIS and Cassini-CIRS}}
  {Jupiter's auroral-related stratospheric heating and chemistry {III}:
  Abundances of {C\textsubscript{2}H\textsubscript{4},
  CH\textsubscript{3}C\textsubscript{2}H, C\textsubscript{4}H\textsubscript{2}
  and C\textsubscript{6}H\textsubscript{6} from Voyager-IRIS and
  Cassini-CIRS}}.{\BBCQ}
\newblock
\APACjournalVolNumPages{Icarus}{328}{}{176--193}.
\PrintBackRefs{\CurrentBib}

\bibitem [\protect \citeauthoryear {%
Sinclair%
\ \protect \BOthers {.}}{%
Sinclair%
\ \protect \BOthers {.}}{%
{\protect \APACyear {2017}}%
}]{%
sinclair17b}
\APACinsertmetastar {%
sinclair17b}%
\begin{APACrefauthors}%
Sinclair, J\BPBI A.%
, Orton, G\BPBI S.%
, Greathouse, T\BPBI K.%
, Fletcher, L\BPBI N.%
, Moses, J\BPBI I.%
, Hue, V.%
\BCBL {}\ \BBA {} Irwin, P\BPBI G\BPBI J.%
\end{APACrefauthors}%
\unskip\
\newblock
\APACrefYearMonthDay{2017}{}{}.
\newblock
{\BBOQ}\APACrefatitle {Jupiter's auroral-related stratospheric heating and
  chemistry {I}: analysis of {Voyager-IRIS} and {Cassini-CIRS} spectra}
  {Jupiter's auroral-related stratospheric heating and chemistry {I}: analysis
  of {Voyager-IRIS} and {Cassini-CIRS} spectra}.{\BBCQ}
\newblock
\APACjournalVolNumPages{Icarus}{292}{}{182--207}.
\PrintBackRefs{\CurrentBib}

\bibitem [\protect \citeauthoryear {%
Sinclair%
\ \protect \BOthers {.}}{%
Sinclair%
\ \protect \BOthers {.}}{%
{\protect \APACyear {2018}}%
}]{%
sinclair18}
\APACinsertmetastar {%
sinclair18}%
\begin{APACrefauthors}%
Sinclair, J\BPBI A.%
, Orton, G\BPBI S.%
, Greathouse, T\BPBI K.%
, Fletcher, L\BPBI N.%
, Moses, J\BPBI I.%
, Hue, V.%
\BCBL {}\ \BBA {} Irwin, P\BPBI G\BPBI J.%
\end{APACrefauthors}%
\unskip\
\newblock
\APACrefYearMonthDay{2018}{}{}.
\newblock
{\BBOQ}\APACrefatitle {Jupiter's auroral-related stratospheric heating and
  chemistry {II}: analysis of {IRTF-TEXES} spectra measured in {December 2014}}
  {Jupiter's auroral-related stratospheric heating and chemistry {II}: analysis
  of {IRTF-TEXES} spectra measured in {December 2014}}.{\BBCQ}
\newblock
\APACjournalVolNumPages{Icarus}{300}{}{305--326}.
\PrintBackRefs{\CurrentBib}

\bibitem [\protect \citeauthoryear {%
Trantham%
}{%
Trantham%
}{%
{\protect \APACyear {2014}}%
}]{%
trantham14}
\APACinsertmetastar {%
trantham14}%
\begin{APACrefauthors}%
Trantham, B.%
\end{APACrefauthors}%
\unskip\
\newblock
\APACrefYearMonthDay{2014}{}{}.
\newblock
\APACrefbtitle {{Juno Jupiter UVS Calibrated Data Archive V1.0}.} {{Juno
  Jupiter UVS Calibrated Data Archive V1.0}.}
\newblock
\APAChowpublished {PDS Atmospheres (ATM) Node}.
\newblock
\begin{APACrefDOI} \doi{10.17189/c32j-7r56} \end{APACrefDOI}
\PrintBackRefs{\CurrentBib}

\bibitem [\protect \citeauthoryear {%
Waite%
\ \protect \BOthers {.}}{%
Waite%
\ \protect \BOthers {.}}{%
{\protect \APACyear {1983}}%
}]{%
waite83}
\APACinsertmetastar {%
waite83}%
\begin{APACrefauthors}%
Waite, J\BPBI H.%
, Cravens, T\BPBI E.%
, Kozyra, J.%
, Nagy, A\BPBI F.%
, Atreya, S\BPBI K.%
\BCBL {}\ \BBA {} Chen, R\BPBI H.%
\end{APACrefauthors}%
\unskip\
\newblock
\APACrefYearMonthDay{1983}{}{}.
\newblock
{\BBOQ}\APACrefatitle {Electron precipitation and related aeronomy of the
  {Jovian} thermosphere and ionosphere} {Electron precipitation and related
  aeronomy of the {Jovian} thermosphere and ionosphere}.{\BBCQ}
\newblock
\APACjournalVolNumPages{Journal of Geophysical Research: Space
  Physics}{88}{A8}{6143--6163}.
\PrintBackRefs{\CurrentBib}

\bibitem [\protect \citeauthoryear {%
West%
\ \protect \BOthers {.}}{%
West%
\ \protect \BOthers {.}}{%
{\protect \APACyear {2004}}%
}]{%
west04}
\APACinsertmetastar {%
west04}%
\begin{APACrefauthors}%
West, R\BPBI A.%
, Baines, K\BPBI H.%
, Friedson, A\BPBI J.%
, Banfield, D.%
, Ragent, B.%
\BCBL {}\ \BBA {} Taylor, F\BPBI W.%
\end{APACrefauthors}%
\unskip\
\newblock
\APACrefYearMonthDay{2004}{}{}.
\newblock
{\BBOQ}\APACrefatitle {Jovian Clouds and Haze} {Jovian clouds and haze}.{\BBCQ}
\newblock
\BIn{} \APACrefbtitle {{Jupiter: The Planet, Satellites and Magnetosphere}}
  {{Jupiter: The Planet, Satellites and Magnetosphere}}\ (\BPGS\ 79--104).
\newblock
\APACaddressPublisher{}{Cambridge University Press}.
\PrintBackRefs{\CurrentBib}

\bibitem [\protect \citeauthoryear {%
Wong%
, Yung%
\BCBL {}\ \BBA {} Friedson%
}{%
Wong%
\ \protect \BOthers {.}}{%
{\protect \APACyear {2003}}%
}]{%
wong03}
\APACinsertmetastar {%
wong03}%
\begin{APACrefauthors}%
Wong, A\BHBI S.%
, Yung, Y\BPBI L.%
\BCBL {}\ \BBA {} Friedson, A\BPBI J.%
\end{APACrefauthors}%
\unskip\
\newblock
\APACrefYearMonthDay{2003}{}{}.
\newblock
{\BBOQ}\APACrefatitle {Benzene and haze formation in the polar atmosphere of
  {Jupiter}} {Benzene and haze formation in the polar atmosphere of
  {Jupiter}}.{\BBCQ}
\newblock
\APACjournalVolNumPages{Geophysical Research Letters}{30}{8}{}.
\PrintBackRefs{\CurrentBib}

\bibitem [\protect \citeauthoryear {%
Woods%
\ \protect \BOthers {.}}{%
Woods%
\ \protect \BOthers {.}}{%
{\protect \APACyear {2009}}%
}]{%
woods09}
\APACinsertmetastar {%
woods09}%
\begin{APACrefauthors}%
Woods, T\BPBI N.%
, Chamberlin, P\BPBI C.%
, Harder, J\BPBI W.%
, Hock, R\BPBI A.%
, Snow, M.%
, Eparvier, F\BPBI G.%
\BDBL {}Richard, E\BPBI C.%
\end{APACrefauthors}%
\unskip\
\newblock
\APACrefYearMonthDay{2009}{}{}.
\newblock
{\BBOQ}\APACrefatitle {Solar irradiance reference spectra {(SIRS)} for the 2008
  whole heliosphere interval {(WHI)}} {Solar irradiance reference spectra
  {(SIRS)} for the 2008 whole heliosphere interval {(WHI)}}.{\BBCQ}
\newblock
\APACjournalVolNumPages{Geophysical Research Letters}{36}{1}{}.
\PrintBackRefs{\CurrentBib}

\bibitem [\protect \citeauthoryear {%
Yao%
\ \protect \BOthers {.}}{%
Yao%
\ \protect \BOthers {.}}{%
{\protect \APACyear {2022}}%
}]{%
yao22}
\APACinsertmetastar {%
yao22}%
\begin{APACrefauthors}%
Yao, Z\BPBI H.%
, Bonfond, B.%
, Grodent, D.%
, Chan{\'e}, E.%
, Dunn, W\BPBI R.%
, Kurth, W\BPBI S.%
\BDBL {}others%
\end{APACrefauthors}%
\unskip\
\newblock
\APACrefYearMonthDay{2022}{}{}.
\newblock
{\BBOQ}\APACrefatitle {On the Relation Between Auroral Morphologies and
  Compression Conditions of {Jupiter's} Magnetopause: Observations From {Juno}
  and the {Hubble Space Telescope}} {On the relation between auroral
  morphologies and compression conditions of {Jupiter's} magnetopause:
  Observations from {Juno} and the {Hubble Space Telescope}}.{\BBCQ}
\newblock
\APACjournalVolNumPages{Journal of Geophysical Research: Space
  Physics}{127}{10}{e2021JA029894}.
\PrintBackRefs{\CurrentBib}

\end{thebibliography}

\end{document}